\documentclass[11pt]{article}%
\usepackage{amssymb}
\usepackage{amsthm}
\usepackage{amsmath}
\usepackage{amsfonts}
\usepackage{graphicx}
\usepackage{hyperref}%
\usepackage[ruled,vlined]{algorithm2e}

\providecommand{\U}[1]{\protect\rule{.1in}{.1in}}
\newtheorem{theorem}{Theorem}
\theoremstyle{plain}

\newtheorem{lemma}{Lemma}

\newtheorem{proposition}{Proposition}
\newtheorem{remark}{Remark}
\newtheorem{assumption}{Assumption}

\newtheorem* {AsmJ}{Assumption J}
\newtheorem* {AsmC}{Assumption C}
\newtheorem*{AsmL}{Assumption L}
\begin{document}


\title{Minimax Risk in Estimating Kink Threshold and Testing Continuity
\thanks{Seo gratefully acknowledges the support from the Ministry of Education of the Republic of Korea and the National Research Foundation of Korea (NRF-2020S1A5A2A03046422) and from the Research Grant of the Center for Distributive Justice at the Institute of Economic Research, Seoul National University. Hidalgo acknowledges financial support from STICERD under the grant "Testing economic shape restrictions’'.}
}
\author{Javier Hidalgo\\London School of Economics
	\and Heejun Lee\\Brown University
\and Jungyoon Lee\\Royal Holloway, University London
\and Myung Hwan Seo\\Seoul National University }

\maketitle

\begin{abstract}
	We derive a risk lower bound in estimating the threshold parameter without knowing whether the threshold regression model is continuous or not. The bound goes to zero as the sample size $ n $ grows only at the cube root rate. Motivated by this finding, we develop a continuity test for the threshold regression model and a bootstrap to compute its \textit{p}-values. The validity of the bootstrap is established, and its finite sample property is explored through Monte Carlo simulations. 
\end{abstract}


\noindent JEL Classification: C12, C13, C24.

\noindent Keywords: Continuity Test, Kink, Risk lower bound, Unknown Threshold.

\thispagestyle{empty}



\section{\textbf{INTRODUCTION}\label{sec:Intro}}
The threshold model has been widely used to model the nonlinearity of time series. For instance, threshold autoregressive  (TAR) model is one of the earliest regime switching models. In its simplest form, it is assumed that there are two regimes. The regime is determined depending on the realization of the threshold variable and the threshold level. See Tong (1990) for a review. Hansen (2000) has extended it to the regression with more Economics application and Hansen and Seo (2002) and Seo (2006) to the threshold cointegration. 
Park and Shintani (2016) and Seo (2008) examined testing issues surrounding threshold effect and unit root. 
Chang \textit{et al.} (2017) proposed an interesting generalization by introducing a latent factor threshold variable, while Lee et al. (2021) extended it further by estimating the factors from an external big data set.

Once one accepts the hypothesis of a threshold effect in the regression function via any of the
available tests, see, e.g., Hansen (1996) and Lee \textit{et al.} (2011) among others,
one is then interested in deciding whether the “segmented” regression model is a model with
a discontinuity (jump) or a model with a kink, since the so-called break/threshold tests are
unable to discriminate between the two models.
A powerful reason to test for a kink comes from the statistical inferential point of view.
As we discuss in Section \ref{sec:model}, the kink design can be represented by a set of restrictions on the
parameter space of the threshold regression model. Thus, the parameters can be consistently
estimated by the unrestricted least squares estimator. Unlike in the linear regression model, where one can make valid inferences based on the unconstrained estimation without knowing
if the constraint holds, inferences in our context have very different statistical properties
under the kink design when using the unrestricted estimation. More specifically, Hidalgo \textit{et al.} (2019) shows that
the rate of convergence of the estimate of the threshold point via unrestricted least squares method is $n^{1/3}$ if there is a kink, which is
in contrast to $n^{1/2}$-rate when the (true) constraint of
a kink is employed in its estimation (Feder, 1975; Chan and Tsay, 1998). If there is not a kink but a jump, then the unrestricted estimate converges in $n$-rate (Chan, 1993), which is also a $\ell_1$-minimax rate (Korostelev, 1987).

On the other hand, we may focus on the fact that the worst-case convergence rate of unrestricted estimate slows down to $n^{1/3}$ if the type of threshold is unknown compared with the situation in which the type of threshold is known so that it is used in the estimation. We show in Section \ref{sec:est_risk} that the cube root convergence rate cannot be improved in terms of  $\ell_1$-risk if the model does not specify the type of threshold. We extend these results to the diminishing threshold model, where the threshold degenerates in polynomial order. The diminishing threshold was introduced by Hansen (2000), and it can be understood as an asymptotic approximation of a small threshold. By allowing the diminishing threshold, we investigate how the size of the threshold affects the performance of estimators. Also, we develop a test valid under both fixed and diminishing threshold effect.

The main contribution of this paper is to develop a testing procedure to distinguish between
jump and kink designs. Hansen (2017) considers inference under the kink design and mentioned
“one could imagine testing the assumption of continuity within the threshold model class. This
is a difficult problem, one to which we are unaware of a solution, and therefore is not pursued in
this paper." We propose a test statistic that is based on the quasi-likelihood ratio and develop
its asymptotic distribution. The difficulty stems from the degeneracy of the hessian matrix of
the expected pseudo-Gaussian likelihood function under the null of continuity. The test is not
asymptotically pivotal since it involves multiple restrictions related to the continuity and conditional heteroscedasticity,
and a bootstrap method is proposed in Section \ref{sec:test_stat_cont} to estimate \textit{p}-values of the test.

We then present the results of a Monte Carlo experiment in Section \ref{sec:MC}, which reports a good finite sample
performance of our bootstrap procedures for the continuity test. In our empirical application in Section \ref{sec:emp}, we employ our test of continuity on the long
span time series data of US real GDP growth and debt-to-GDP ratio data used in Hansen
(2017) which had fitted the kink model. Our test of continuity rejects the null of continuity, and
we present the estimated jump model.
We also consider data from Sweden and find substantial variations across
countries not only in the values of parameter estimates but also in the results of tests on the
presence of threshold effect and continuity.

\section{\textbf{MODEL AND ASSUMPTIONS}\label{sec:model}}

We consider a threshold/segmented regression model
\begin{equation}
Y_i=X_{i}^{\prime}\beta+X_{i}^{\prime}\delta\mathbb{I}\left\{  Q_{i}%
>\tau\right\}  +U_{i}\text{,}\label{eq:model}%
\end{equation}
where $\mathbb{I}\left\{  \cdot\right\}  $ denotes the indicator function, $Y_{i}$ is dependent variable and
$X_{i}$ is a $d$-dimensional vector of regressors. The parameter $\tau$ represents a change/break-point or threshold, taking values in a compact
parameter space $\mathbb{T}$ which lies in the interior of the domain of
the threshold variable $Q_{i}$. In addition, we assume that
$\delta\neq0$, which implies that the model has a threshold effect.

As mentioned before, we consider the case where the conditional expectation of $Y_{i}$ given the regressor $X_{i}$ is allowed to be either continuous, i.e., to have a kink, or discontinuous, i.e., to have a jump. 
We let the threshold variable $Q_{i}$ be an element of the covariate vector $X_{i}$ since otherwise, it would not be possible for the regression function to be continuous. 
We shall decompose the $d$-dimensional parameters and regressors as follows:
\begin{equation}
X_{i}=\left(  1,X_{i2}^{\prime},Q_{i}\right)  ^{\prime};\ \ \ \delta=\left(
\delta_{1},\delta_{2}^{\prime},\delta_{3}\right)  ^{\prime}\text{,}%
\label{x_not}%
\end{equation}
where $\delta$ is partitioned to match the dimensionality of $X_{i}$ and $X_{i2}$ is a $(d-2)$-dimensional vector. Also we
shall abbreviate $\mathbb{I}_{i}\left(  \tau\right)  =\mathbb{I}\left\{
Q_{i}>\tau\right\}  $ and $X_{i}\left(  \tau\right)  =\left(
X_{i}^{\prime},X_{i}^{\prime}\mathbb{I}_{i}\left(  \tau\right)  \right)
^{\prime}$, so that we can write $\left(  \ref{eq:model}\right)  $ as
\begin{align}
Y_{i}  & =X_{i}^{\prime}\beta+\delta_{1}\mathbb{I}_{i}\left(  \tau\right)
+X_{i2}^{\prime}\delta_{2}\mathbb{I}_{i}\left(  \tau\right)  +\delta
_{3}Q_{i}\mathbb{I}_{i}\left(  \tau\right)  +U_{i}\label{jd}\\
& =X_{i}\left(  \tau\right)^{\prime}\alpha +U_{i}\text{,}%
\quad\text{where}\quad\alpha=(\beta^{\prime},\delta^{\prime})^{\prime}%
\text{.}\label{model_ab}%
\end{align}

\noindent \textbf{Notation}. 
Before stating some regularity assumptions on the model, we introduce
some extra notations. Let $f\left(  \cdot\right)  $ denote the density function
of $Q_{i}$ and $\sigma^{2}\left(  \tau\right)  =E\left(  U_{i}%
^{2}\mid Q_{i}=\tau\right)  $, the conditional variance function of the error term, while $\sigma^{2}=E(U_{i}^{2})$ denotes the unconditional
variance. Denote $d\times d$ matrices $D\left(  \tau\right)  =E\left(
X_{i}X_{i}^{\prime}|Q_{i}=\tau\right)  $, $V\left(  \tau\right)  =E\left(
X_{i}X_{i}^{\prime}U_{i}^{2}|Q_{i}=\tau\right)  $ and let
$D=D\left(  \tau_{0}\right)  $ and $V=V\left(  \tau_{0}\right)  $. As
usual the \textquotedblleft$0$\textquotedblright\ subscript on a parameter
indicates its true unknown value. Finally, let $M=E(\mathbf{X}_{i}\mathbf{X}_{i}%
^{\prime})$ and $\Omega=E(\mathbf{X}_{i}\mathbf{X}_{i}^{\prime}U_{i}^{2})$
with $\mathbf{X}_{i}=X_{i}\left(  \tau_{0}\right)  $.
\newline

We shall now introduce some regularity conditions. 

\begin{assumption} \label{A:Z}
Let $\left\{  X_{i},U_{i}\right\}  _{i\in\mathbb{Z}}$ be a strictly
stationary, ergodic sequence of random variables such that their $\rho$-mixing
coefficients satisfy $\sum_{m=1}^{\infty}\rho_{m}^{1/2}<\infty$ and $E\left(
U_{i}|{{\mathcal{F}}}_{i-1}\right)  =0$, where ${{\mathcal{F}}}_{i}$
is the filtration up to time $i$. Furthermore, $M,\Omega>0$, $E\left\Vert
X_{i}\right\Vert ^{4}<\infty$, $E\left\Vert X_{i}U_{i}\right\Vert
^{4}<\infty$ and $E\left\vert U_{i}\right\vert ^{4+\eta}<\infty$ for
some $\eta>0$.
\end{assumption}

\begin{assumption} \label{A:Q}
The functions $f\left(  \tau\right)  $, $V\left(  \tau\right)  \ $and
$D\left(  \tau\right)  $ are continuous at $\tau=\tau_{0}$. For all
$\tau\in\mathbb{T}$, the functions $f\left(  \tau\right)  $, $E\big(X_{i}%
X_{i}^{\prime}\mathbb{I}\left\{  Q_{i}\leq\tau\right\}  \big)$ and $Var \left(
X_{i2}|Q_{i}=\tau\right)  $ are positive and continuous, and
the functions $f\left(  \tau\right)  $,$\ E\big(|X_{i}|^{4}|Q_{i}%
=\tau\big)$ and $E\big(|X_{i}U_{i}|^{4}|Q_{i}=\tau\big)$ are
bounded by some $C<\infty$.
\end{assumption}

These are similar to those in Hansen (2000). Note that the SETAR
model of Tong $\left(  1990\right)  $ satisfies Assumption \ref{A:Z}. The condition for the conditional moment $Var\left(  X_{i2}|Q_{i}=\tau\right)  $ is
written in terms of $X_{i2}$ as the other elements in $X_{i}$ are fixed given
$Q_{i}=\tau$. While we allow conditional heteroscedasticity of a general
form, Assumption \ref{A:Q} requires continuity of the conditional variance
function $\sigma^{2}(\cdot)$ at $\tau_{0}$. We need to estimate the conditional variance via nonparametric methods.

We shall emphasize that the model $(\ref{eq:model})$ encompasses both the kink and jump models. The kink model is characterized by the continuity restriction:%
\begin{AsmC}
$\delta_{30}\neq0$ and
\begin{equation}
\delta_{10}+\delta_{30}\tau_{0}=0;\ \ \ \delta_{20}=0\text{.}%
\label{eq:conti}%
\end{equation}
\end{AsmC}
\noindent Note that we require $\delta_{30}$ to be nonzero to identify $\tau_0$. Under $\left(
\ref{eq:conti}\right)  $, we observe that $\left(  \ref{jd}\right)  $ becomes%
\begin{equation}
Y_{i}=X_{i}^{\prime}\beta_{0}+\delta_{30}(Q_{i}-\tau_{0})\mathbb{I}%
_{i}\left(  \tau_{0}\right)  +U_{i}\text{.}\label{eq:simp}%
\end{equation}

For the sake of completeness, we define the jump threshold:
\begin{AsmJ}
$\delta_{0}\neq0$ and
\begin{equation}
\delta_{0}'D\delta_{0}>0 \text{.}%
\label{eq:jump}%
\end{equation}
\end{AsmJ}
In the following sections, we allow for the threshold effect $ \delta_0 $ to converge to zero at a  polynomial rate, as in Hansen (2000). Specifically, $\delta_0 = d_0 \cdot n^{-\varphi}$ where $\varphi \geq 0$ and $d_0$ is fixed over $n$. We call the case where $\varphi = 0$ a fixed threshold and the case where $\varphi > 0$ a diminishing threshold.

\section{\textbf{ESTIMATORS AND RISK BOUND}\label{sec:est_risk}}

This section elaborates on how the continuity restriction affects the estimation of the threshold location $\tau_0$. As mentioned before, when the continuity restriction is not employed in the estimation, the rate of convergence is either $n$ if there is a jump or $n^{1/3}$ if there is a kink, which means that the worst-case performance of the unrestricted estimator is $n^{1/3}$ under the situation that the type of threshold is unknown. A generalized result that includes the diminishing threshold effect is presented in Proposition \ref{Consistency}. One may pursue to propose an estimation procedure that outperforms the unrestricted estimator with respect to the worst-case convergence rate. However, it is impossible to overcome the cube-root rate in $\ell_1$-minimax sense if the information about the threshold type is unavailable, as we show in Proposition \ref{Lower Bound}.

\subsection{Estimators\label{sec:LSE}}

We choose the residual sum of squares as the objective function. Denote parameters by $\theta=\left(  \alpha^{\prime},\tau\right)  ^{\prime}\in\mathbb{R}^{2d+1}$ and denote the objective function by  $\mathbb{S}_n$ where \begin{equation}
{{\mathbb{S}}}_{n}\left(  \theta\right)  =\frac{1}{n}\sum_{i=1}^{n}\left(
Y_{i}-X_{i}^{\prime}(\tau)\alpha  \right)  ^{2}\text{.}%
\label{s_theta}%
\end{equation}
If the continuity restriction is not imposed on the true parameter $\theta_0$, then it can be estimated by minimizing the objective function, that is,
\begin{equation}
\widehat{\theta}=\left(  \widehat{\alpha}^{\prime},\widehat{\tau}\right)
^{\prime}=\underset{\theta\in\Theta}{\operatorname{argmin}}\,\mathbb{S}%
_{n}\left(  \theta\right)  \text{,}\label{theta_hat}%
\end{equation}
where $\Theta=\Lambda \times \mathbb{T}$ is a compact set in $ \mathbb{R}^{2d+1}$. Following convention, we let $ \widehat{\tau} $ be an element of $ \{Q_i\} $.

On the other hand, we can minimize $\left(  \ref{s_theta}\right)  $ among the elements of $\Theta$ that satisfy 
constraints in $\left(  \ref{eq:conti}\right)$, yielding the constrained least
squares estimator (CLSE):
\begin{equation}
\widetilde{\theta}=\left(  \widetilde{\alpha}^{\prime},\widetilde{\tau
}\right)  ^{\prime}:=\underset{\theta\in\Theta:\delta_{1}+\delta_{3}%
	\tau=0;\delta_{2}=0}{\operatorname{argmin}}\mathbb{S}_{n}\left(
\theta\right)  \text{.}\label{theta_tilde}%
\end{equation}

Since criterion is not smooth, we compute the unconstrained least squares estimator (LSE) as a two-step algorithm. Since the criterion $ \mathbb{S}_n $ is in fact a step function along $ \tau $ with jumps at each $ Q_i $, we may first discretize the parameter space of threshold $\mathbb{T}$ as $\mathbb{T}_{n}=\mathbb{T}\cap\left\{  Q_{1},...,Q_{n}\right\}  $
to find $\widehat{\tau}$. 
Then, find $\widehat{\alpha}(\tau)$ which minimizes the sum of squared errors for each $\tau$:
\begin{equation}
\widehat{\alpha}\left(  \tau\right)  =\underset{\alpha\in\Lambda
}{\operatorname{argmin}}\text{~}\frac{1}{n}\sum_{i=1}^{n}\left(  Y_{i}%
-X_{i}^{\prime}\left(  \tau\right)  \alpha  \right)^{2}\label{s_alpha}%
\end{equation}
Finally, we define the least square estimator $\widehat{\tau}$ as the minimizer of the sum of squared errors:
\begin{equation}
\widehat{\tau}=\underset{\tau\in\mathbb{T}_{n}}{\operatorname{argmin}%
}\,\widehat{\mathbb{S}}_{n}\left(  \tau\right)  \text{,}\label{s_gamma}%
\end{equation}
where
\begin{equation}
\widehat{\mathbb{S}}_{n}\left(  \tau\right)  =\frac{1}{n}\sum_{i=1}%
^{n}\left(  Y_i-X_i^{\prime}\left(  \tau\right)  \widehat{\alpha}\left(
\tau\right)  \right)  ^{2}\text{.}\label{ssngm}%
\end{equation}
Then, our estimator of $\alpha$
is $\widehat{\alpha}=\widehat{\alpha}\left(  \widehat{\tau}\right)$. The constrained least squares estimator can be obtained similarly.

Suppose that $\delta_0 = d_0\cdot n^{-\varphi}$, for $0\leq\varphi<1/2$ and a nonzero vector $d_0\in\mathbb{R}^d$. When $\varphi=0$, the rate of convergence is $n$ if there is a jump (Chan, 1993), whereas it is only $n^{1/3}$ if there is a kink and the restriction is not used in the estimation (Hidalgo \textit{et al.}, 2019). If the restriction is used in the estimation, the rate of convergence is $n^{1/2}$. On the other hand, when $\varphi>0$, the rate is $n^{1-2\varphi}$ when there is a jump (Hansen, 2000), and we shall show that when there is a kink, the rate of convergence becomes $n^{(1-2\varphi)/{3}}$ if the restriction is not used in the estimation.

\begin{proposition}
\label{Consistency}\textit{Let Assumptions \textbf{C,} \textit{\ref{A:Z}},
and \ref{A:Q} hold. If  $\delta_0 = d_0\cdot n^{-\varphi}$ for some $0\leq\varphi<1/2$ and $ d_0 \neq 0 $, we have that}
\[
\widehat{\alpha}-\alpha_{0}=O_{p}\big(n^{-1/2}\big)\ \ \ \text{and}%
\ \ \ \ \widehat{\tau}-\tau_{0}=O_{p}\big(n^{(2\varphi-1)/3}\big)\text{.}%
\]

\end{proposition}
This proposition generalizes the rate of convergence of
the LSE $\widehat{\theta}$ under the fixed threshold assumption explored in Hidalgo \textit{et al.} (2019) to encompass the diminishing threshold. 

\subsection{Risk Bound}
In this section we shall develop an $\ell_1$-minimax lower bound in estimating the threshold $\tau$. The $\ell_1$-risk of an estimator $\widehat{\tau}$ for $\tau$ is defined as
\begin{equation}
    \mathcal{R}_n(\widehat{\tau}\text{ ; }\alpha, \tau, \mathbb{Q}_n) = E\bigl(|\widehat{\tau}-\tau|\bigr),
\end{equation}
where the expectation depends on $\alpha_0$, $\tau_0$, and $\mathbb{Q}_n$, the joint distribution of $\{X_{i},U_{i}\}_{i=1}^n$. Let $\mathcal{P}(n,\kappa, \underline{\sigma}^2, \overline{f})$ denote the class of joint distributions of $\{Y_i,X_i\}_{i=1}^n$ such that $Y_i = X_i(\tau)^{\prime}\alpha+U_{i}$ for all $i\in\mathbb{Z}$, $|\delta_3|\geq\kappa$, and $f(\tau)\leq \overline{f}, \sigma^2(\tau)\geq\underline{\sigma}^2$ for all $\tau\in\mathbb{T}$. We evaluate the performance of an estimator based on the most adverse choice of the distribution $\mathbb{P}_n\in\mathcal{P}(n,\kappa, \underline{\sigma}^2, \overline{f})$, namely,
\begin{equation}
    \sup_{\mathbb{P}_n\in\mathcal{P}(n,\kappa, \underline{\sigma}^2, \overline{f})}\mathcal{R}_n(\widehat{\tau}\text{ ; }\alpha(\mathbb{P}_n), \tau(\mathbb{P}_n), \mathbb{Q}_n(\mathbb{P}_n))\text{,}\label{min}
\end{equation}
where $\alpha(\mathbb{P}_n),\tau(\mathbb{P}_n)$, and $\mathbb{Q}_n(\mathbb{P}_n)$ make the joint distribution of $\{Y_i, X_i\}_{i=1}^n$ equal to $\mathbb{P}_n$. We will show that the worst-case risk $\left(\ref{min}\right)$ of any estimator cannot tend to zero faster than the cube-root rate by providing a lower bound for the $\ell_1$-minimax risk.

Our lower bound is valid even for a restrictive subclass of $\mathcal{P}(n,\kappa, \underline{\sigma}^2, \overline{f})$ induced by Assumption \ref{A:Z}, that is,
\begin{AsmL}
Let $\{X_i,U_i\}_{i\in\mathbb{Z}}$ be a sequence of independent and identically distributed random vectors. Assume that  $U_i$ follows $\mathcal{N}(0,\sigma^2(Q_i))$ given $X_i$. 
\end{AsmL}
Even if we assume that the $\delta_2$ is known to be zero, the cube-root lower bound cannot be improved. Let $\eta$ be the diameter of $\mathbb{T}$, that is, $\eta = \sup_{\tau_1,\tau_2\in\mathbb{T}}|\tau_2 - \tau_1|$. For notational convenience, we focus on $\mathbb{T}\subset (0,1)$ since for any interval $(a,b)$, there exists a trivial affine transformation to $(0,1)$. Let $\kappa = \kappa_0 n^{-\varphi}$. If $\varphi>0$, it represents the diminishing threshold effect. Then the minimax risk is lower bounded as follows:
\begin{proposition}
\label{Lower Bound}\textit{Assume that} $\mathbb{T}$ is a closed interval in $(0, 1)$ and $\delta_2=0$. \textit{Under Assumption \textbf{L,} we have that}
 \[
 \inf_{\widehat{\tau}}\sup_{\mathbb{P}_n\in \mathcal{P}(n,\kappa,\underline{\sigma}^2, \overline{f})} E_{\mathbb{P}_n} \bigl(|\widehat{\tau}-\tau(\mathbb{P}_n)|\bigr) \geq \begin{cases}
   \frac{\underline{\sigma}^{2/3}}{3\overline{f}^{1/3}\kappa_0^{2/3}}n^{(2\varphi-1)/3} & \text{if $n^{(1-2\varphi)}\geq \frac{3\underline{\sigma}^2}{\overline{f}\kappa^2\eta^3}$}\\
   \frac{1}{4}\eta & \text{if $n^{(1-2\varphi)}< \frac{3\underline{\sigma}^2}{\overline{f}\kappa^2\eta^3}$} 
 \end{cases},
 \]
for $0\leq\varphi<1/2$, where the infimum is taken over all estimators $\widehat{\tau}$ of $\tau$.  
\end{proposition}

Note that there are reasonable relationships between the constant factor multiplied to $n^{-1/3}$ and nuisance parameters in Proposition \ref{Lower Bound}. When the noise $\underline{\sigma}^2$ is a large constant or the minimal slope change $\kappa$ is small, the estimation of $\tau$ becomes harder. 

Thus far, in this section, we considered one of the simplest forms of the threshold model except that it includes both the jump and kink threshold. Therefore, the major complexity that causes the slow decay rate of the minimax risk lies in the fact that it is unknown whether the regression function is continuous or not. From this observation, we can see that there would be little gain from searching for an estimator with better accuracy without knowing the continuity of regression function, which motivates the test for the continuity. 

\begin{remark}
We derived the risk lower bound under Assumption \textbf{L} instead of Assumption 1. As mentioned earlier, Assumption \textbf{L} is more restrictive than Assumption 1. In some sense, Assumption \textbf{L} is a favorable scenario of Assumption 1. Since the worst-case performance under the favorable scenario cannot be better than that under the general scenario, the risk bound in Proposition \ref{Lower Bound} is also valid under Assumption 1. The implication of Proposition 2 is that the minimax risk cannot tend to zero faster than the cube-root rate even under the favorable scenario if the type of the threshold is unknown.
\end{remark}

\section{\textbf{TESTING CONTINUITY}\label{sec:test_stat_cont}}

This section considers testing of the continuity restriction, stated
formally as
\begin{equation}
H_{0}:\delta_{10}+\delta_{30}\tau_{0}=0\quad\text{and}\quad\delta
_{20}=0\text{,}\label{H0}%
\end{equation}
along with an auxiliary condition of $\delta_{30}\neq0$ to ensure the
identification of the threshold point $\tau_{0}$.

The alternative hypothesis is its negation
\begin{equation}
H_{1}:\delta_{10}+\delta_{30}\tau_{0}\neq0\quad\text{and/or}\quad\delta
_{20}\neq0\text{.}\label{e:h1}%
\end{equation}
Provided that $Var\left[  X_{i2}|Q_{i}=\tau_{0}\right]  >0$,
the hypothesis $H_{1}$ yields that%
\[
E\left[  \left(  \delta_{10}+\delta_{30}\tau_{0}+%
X_{i2}^{\prime}\delta_{20}\right)  ^{2}|Q_{i}=\tau_{0}\right]  >0\text{,}%
\]
which implies that the regression function has a jump (non-zero change) at
$Q_{i}=\tau_{0}$ with positive probability. As mentioned in the previous section, we develop a test valid for both fixed and diminishing threshold. In order to obtain such a test, we extend the earlier results of Hidalgo \textit{et al.} (2019) about the fixed threshold to the diminishing threshold.

\subsection{Continuity Test}

To develop the test, we first need to derive the asymptotic distributions of the LSE $\widehat{\theta}$ and CLSE $\widetilde{\theta}$ under Assumption \textbf{C}. Feder (1975) and later Chan and Tsay (1998) or Hansen (2017) have already established the asymptotic normality of $\widetilde{\theta}$ with the standard squared root consistency. Thus, we only need to examine the asymptotic properties of $\widehat{\theta}$. We present the asymptotic distribution of $\widehat{\theta}$ under the null.

\begin{theorem}
\label{Th:AD_cube} Let Assumptions \textit{\textbf{C,} \textit{\ref{A:Z}}, and
\ref{A:Q}} hold, and $B_{1}\left(  \cdot\right)  $, $B_{2}\left(
\cdot\right)  $ be two independent standard Brownian motions. Define $W\left(
g\right)  :=B_{1}\left(  -g\right)  \mathbb{I}\left\{  g<0\right\}
+B_{2}\left(  g\right)  \mathbb{I}\left\{  g>0\right\}  $. Let $\delta_0 = d_0\cdot n^{-\varphi}$. If $0\leq \varphi <1/2$,
\begin{align*}
& n^{1/2}(\widehat{\alpha}-\alpha_{0})\overset{d}{\longrightarrow}%
\mathcal{N}\left(  0,M^{-1}\Omega M^{-1}\right) \\
& n^{(1-2\varphi)/3}(\widehat{\tau}-\tau_{0})\overset{d}{\longrightarrow}%
\underset{g\in\mathbb{R}}{\operatorname{argmax}}\big(2d_{30}\sqrt
{\frac{\sigma^{2}\left(  \tau_{0}\right)  f\left(  \tau_{0}\right)  }{3}%
}W\left(  g^{3}\right)  +\frac{d_{30}^{2}}{3}f\left(  \tau_{0}\right)
\left\vert g\right\vert ^{3}\big)\text{,}%
\end{align*}
where the two limit distributions are independent of each other.
\end{theorem}
This result is an extension of Theorem 1 in Hidalgo \textit{et al.} (2019) where only the fixed threshold case, $\varphi=0$, is considered.
\subsection{Test Statistic}

Our testing problem is non-standard. First, the score-type test is not
straightforward due to the non-differentiability of the criterion function
$\mathbb{S}_{n}\left(  \theta\right)  $ with respect to $\tau$. Second, the
unconstrained estimators $\widehat{\tau}$ and $\widehat{\delta}$ converge at
different rates to different family of probability distribution functions
making the construction of a Wald-type test non-obvious. Thus, we consider a
quasi-likelihood ratio statistic, which compares the constrained sum of
squared residuals with the unconstrained one, i.e.
\begin{equation}
T_{n}=n\frac{\widetilde{\mathbb{S}}_{n}-\widehat{\mathbb{S}}_{n}}%
{\widehat{\mathbb{S}}_{n}}\label{unscaled}%
\end{equation}
where $\widehat{\mathbb{S}}_{n}=\mathbb{S}_{n}\left(  \widehat{\mathbb{\theta
}}\right)  $ and $\widetilde{\mathbb{S}}_{n}=\mathbb{S}_{n}\left(
\widetilde{\theta}\right)  $.

Deriving the asymptotic distribution of $T_{n}$ is also non-standard due to
the lack of expansion of the criterion function $\mathbb{S}_{n}\left(
\theta\right)  $ with respect to $\tau$. Therefore, we employ the approach
developed by Lee \textit{et al}. $\left(  2011\right)  $, which reformulates
the statistic as a continuous functional of a stochastic process over an
expanded domain. In particular, denote by $I_{d}$ and $0_{a\times b}$ the
identity matrix of dimension $d$ and the matrix of zeros of dimension $a\times
b$, respectively, and let
\[
R=\left(
\begin{array}
[c]{cc}%
I_{d} & 0_{d\times d}\\
0_{1\times d} & -\tau_{0}:0_{1\times\left(  d-2\right)  }:1\\
0_{1\times d} & -\beta_{30}-\delta_{30}:0_{1\times\left(  d-1\right)  }%
\end{array}
\right)  \text{.}%
\]
Define a Gaussian process
\begin{align*}
\mathbb{K}\left(  h,g,\ell\right)   & =\ell^{\prime}E\left(R\mathbf{X}_{i}%
\mathbf{X}_{i}^{\prime}R^{\prime}\right)\ell+h^{\prime}E\left(\mathbf{X}_{i}\mathbf{X}%
_{i}^{\prime}\right)h-2\left(  \ell^{\prime}R+h^{\prime}\right)  B\\
& +\big(2d_{30}\sqrt{\frac{\sigma^{2}\left(  \tau_{0}\right)  f\left(
\tau_{0}\right)  }{3}}W\left(  g^{3}\right)  +\frac{d_{30}^{2}}%
{3}f\left(  \tau_{0}\right)  \left\vert g\right\vert ^{3}\big),
\end{align*}
where $B$ follows $\mathcal{N}\left(  0,\Omega\right)  $ and is independent of
the Gaussian process $W\ $that was introduced in Theorem \ref{Th:AD_cube}.

\begin{theorem}
\label{Th:Q_n}Under Assumptions\textit{\ \textit{\ref{A:Z}} and \ref{A:Q},}
and \textit{\textbf{C}} with $\delta_{30}=d_{30}\cdot n^{-\varphi}%
,~0\leq\varphi<1/2,\ $ and $d_{30}\neq0,$
\[
T_{n}\overset{d}{\longrightarrow}\left(\min_{\ell:g=0,h=0}\mathbb{K}\left(  h,g,\ell\right) -\min_{h:g=0,\ell=0}\mathbb{K}\left(
h,g,\ell\right)  -\min_{g:h=0,\ell=0}\mathbb{K}\left(  h,g,\ell\right)
 \right)/\sigma^{2}\text{.}%
\]

\end{theorem}

It is worthwhile to note that $\delta_{30}$ is allowed to degenerate at the
rate $n^{-\varphi}$ as well as stay fixed when $\varphi=0$. We allow non-zero
$\varphi$ to examine the property of the continuity test when $\delta_{30}$ is
small, and thus the identification of $\tau_{0}$ is relatively weak. Along
with the Monte Carlo experiments reported in Section \ref{sec:MC}, this
theorem provides support for the good finite-sample performance of our
continuity test based on the statistic $T_{n}$ even when $\delta_{30}$ is small.

Next, we remark on the auxiliary assumption that $\delta_{30}\neq0.$ Recall
that the discussion following (\ref{eq:conti}) that without $\delta_{30}%
\neq0\ $the continuity restriction (\ref{eq:conti}) implies that $\delta
_{0}=0$ and thus, the null model is not a model with a kink but a linear
regression model, a consequence being that $\tau_{0}$ is unidentifiable as well. Indeed,
testing that $\delta_{0}=0$ is a classic non-standard testing problem, also
known as Davies' problem, where the null hypothesis induces a loss of
identification. It has been studied intensively in the literature as in e.g.
Hansen $\left(  1996\right)  $ and Lee \textit{et al}. $\left(  2011\right)  $
to cite a few. Our testing problem is different from this Davies' problem and
does not involve a loss of identification. Another related testing problem is
the testing of the jump hypothesis against more general transition functions
like Kim and Seo $\left(  2017\right)  $.

Next, we establish the consistency of the test. Since $\widehat{\mathbb{S}%
}_{n}\overset{p}{\longrightarrow}EU_i^{2}$ while $\widetilde
{\mathbb{S}}_{n}\overset{p}{\longrightarrow}EU_i^{2}+c$ for some
$c>0,$ which is due to the rank conditions in Assumptions \ref{A:Z} and
\ref{A:Q}, $T_{n}$ diverges to $+\infty$ under the alternative. Formally,

\begin{theorem}
\label{Th:Consistency of Qn} Under Assumptions \ref{A:Z} and \ref{A:Q} and
the alternative hypothesis $H_{1}$ (\ref{e:h1}),
\[
P\left\{  T_{n}>c\right\}  \rightarrow1\text{,}%
\]
for any $c<\infty.$
\end{theorem}

As the limiting distribution of $T_{n}$ is not pivotal as it depends on the multiple
restrictions and conditional heteroskedasticity, it is not practically useful to derive an explicit expression of its limit distribution, and
hence we do not pursue it here. Instead, to compute its critical values, we proceed by examining a valid bootstrap to estimate the \textit{p}-values of the test statistic.

\subsection{\textbf{Bootstrapping Continuity Test}\label{sec:btrp c_test}}

This section provides a bootstrap procedure for the test of continuity based
on the $T_{n}$ statistic. We shall mention that the bootstrap-based test inversion
confidence interval for the unknown threshold parameter $\tau_{0}$ is developed in Hidalgo \textit{et al.} (2019).
We proceed as follows:

\begin{algorithm}
\SetAlgoLined
 \textbf{\emph{STEP 1}} Obtain both LSE $\widehat{\theta}=\left(  \widehat
{\alpha}^{\prime},\widehat{\tau}\right)  ^{\prime}$ and CLSE $\widetilde
{\theta}=\left(  \widetilde{\alpha}^{\prime},\widetilde{\tau}\right)
^{\prime}$ of $\theta_{0}=\left(  \alpha_{0}^{\prime},\tau_{0}\right)
^{\prime}$ as given in (\ref{s_alpha}), (\ref{s_gamma}) and (\ref{theta_tilde}%
), and compute the least squares residuals
\[
\widehat{U}_i=Y_{i}-X_{i}\left(
\widehat{\tau}\right)^{\prime}\widehat{\alpha} ,\text{ \ \ }i=1,...,n\text{.}%
\]

\textbf{\emph{STEP 2}} Generate $\left\{  \eta_{i}\right\}  _{i=1}^{n}$ as
independent and identically distributed zero mean random variables with unit
variance and finite fourth moments, and compute%
\[
Y_{i}^{\ast}=X_{i}\left(
\widetilde{\tau}\right)^{\prime}\widetilde{\alpha}  +\widehat{U}_{i}\eta_{i}\text{,\ \ \ }i=1,...,n\text{.}%
\]

\textbf{\emph{STEP 3}} Using $\{Y_{i}^{\ast}\}_{i=1}^{n}$ and $\{X_{i}%
\}_{i=1}^{n},$ construct the bootstrap statistic $T_{n}^{\ast}$ as in
(\ref{unscaled}) of Section \ref{sec:test_stat_cont}. Specifically,
\[
T_{n}^{\ast}=n\frac{\widetilde{\mathbb{S}}_{n}^{\ast}-\widehat{\mathbb{S}}%
_{n}^{\ast}}{\widehat{\mathbb{S}}_{n}^{\ast}}\text{,}%
\]
where
\begin{align*}
\widehat{\mathbb{S}}_{n}^{\ast}  & =\min_{\theta}\frac{1}{n}\sum_{i=1}%
^{n}\left(  Y_{i}^{\ast}- X_{i}\left(
\tau\right)^{\prime} \alpha\left(  \tau\right)  \right)  ^{2},\\
\widetilde{\mathbb{S}}_{n}^{\ast}  & =\min_{\theta:\delta_{1}+\delta_{3}%
\tau_{0}=0;\delta_{2}=0}\frac{1}{n}\sum_{i=1}^{n}\left(  Y_{i}^{\ast}%
-  X_{i}\left(  \tau\right)^{\prime} \alpha\left(  \tau\right) \right)
^{2}\text{.}%
\end{align*}

\textbf{\emph{STEP 4}} Compute the bootstrap $p$-value, $p^{\ast}$ by
repeating\emph{\ STEPS\ 2-3} $B$ times and obtain the proportion of times that
$T_{n}^{\ast}$ exceeds the sample statistic $T_{n}$ given in (\ref{unscaled}).
\caption{Bootstrapping the Continuity Test}
\end{algorithm}

\newpage

The validity of this procedure is given in the following theorem. As usual,
the superscript \textquotedblleft$^{\ast}$\textquotedblright{\Large \ }%
indicates the bootstrap quantities and convergences of bootstrap statistics
conditional on the original data. The notation $\overset{d^{\ast}%
}{\longrightarrow}$ \textit{in Probability }signifies the the convergence in
probability of the random distribution functions of the bootstrap statistics
in terms of the uniform metric.

\begin{theorem}
\label{Th:QBoot}Suppose Assumptions \ref{A:Z} and \ref{A:Q} hold. Then,
under Assumption \textbf{C},
\[
T_{n}^{\ast}\overset{d^{\ast}}{\longrightarrow}T\ \ \ \ \text{in Probability,}%
\]
where $T$ denotes the limit variable in Theorem \ref{Th:Q_n}.
\end{theorem}

\section{\textbf{Monte Carlo Experiment}}\label{sec:MC}

As in Hidalgo et al. (2019, Section 5), our simulation is based on the following three specifications:
\begin{align*}
&  A:\,Y_{i}=2+3Q_{i}+\delta Q_{i}\mathbb{I}\left\{  Q_{i}>\tau_{0}\right\}
+U_i,\ \,\tau_{0}=E(Q_i)=0,\\
&  B:\,Y_{i}=2+3Q_{i}+\delta Q_{i}\mathbb{I}\left\{  Q_{i}>\tau_{0}\right\}
+U_i,\ \,\tau_{0}=E(Q_i)=2,\\
&  C:\,Y_{i}=2+3X_{i}+\delta X_{i}\mathbb{I}\left\{  Q_{i}>\tau_{0}\right\}
+U_i,\ \,\tau_{0}=E(Q_{i})=2 .
\end{align*}
Settings B and C are jump models considered in Hansen (2000, Section 4.2), and
setting A represents the kink case.
However, our data generating process differs from Hansen (2000)
in that we assume the conditional heteroscedasticity in $U_i$ such that $U_i=|Q_i|e_{i}$ where $\left\{  e_{i}\right\}  _{i\geq1}$ and $\left\{
Q_{i}\right\}  _{i\geq1}$ were generated as mutually independent and
independent and identically distributed ($i.i.d.$) normal random variables
with unit variance. This leads to conditional heteroscedasticity of the form
$E(U_i^{2}|Q_i)=Q_i^{2}$, in contrast to Hansen (2000) where
$U_i$ was generated from $\mathcal{N}(0,1)$. In A, we generated $X_{i}$ as
$i.i.d.$ draws from $\mathcal{N}(2,1)$, independent of $\left\{  U_{i}\right\}
_{i\geq1}$ and $\left\{  Q_i\right\}  _{i\geq1}$. For the grid $\mathbb{T}_{n} $
used in estimation of $\tau_{0}$, we discard $10\%$ of extreme values of
realized $Q_{i}$ and use $n/2$ number of equidistant points.

We investigate finite-sample performance of the bootstrap-based test of
continuity proposed in Section \ref{sec:test_stat_cont}. Results are based on
10,000 iterations, with one bootstrap per iteration, using the warp-speed
method of Giacomini, Dimitris and White $(2013)$. Table 1 presents Monte Carlo
size results of the test for nominal size $s=0.1,0.05,0.01$. We first try two
settings for $\delta$ that are in line with conditions of Theorem 2: for
columns 2-4 in rows 3-6, $\delta$ is fixed at 2\footnote{$\delta=2$ was the
largest value of $\delta$ tried in Hansen (2000), although that paper only
looks at inference on $\tau$ in jump setups B and C.}, while $\delta$ is
shrinking in $n$ for columns 5-7, with $\delta=n^{-1/4}\sqrt{10}/4 = 0.25,
0.1988, 0.1672, 0.1406 $ for $n=100, 250, 500, 1000$, respectively.
$\delta=0.25$ was the smallest $\delta$ used in Hansen (2000) and by letting
it diminish further for $n=250, 500, 1000$, we hope to investigate size
performance of our test for very small $\delta$. The results show satisfactory
size performance for both cases, with the fixed $\delta$ case producing better
size results, as expected. It is reassuring that the size performance is
satisfactory for $\delta$ as small as $0.1406$ in the diminishing $\delta$
case. We have also tried $\delta=n^{-1/2}\sqrt{10}/4 = 0.0791, 0.05, 0.0354,
0.025 $ for $n=100, 250, 500, 1000$ and $\delta=0$. These settings are outside
the scope of the current paper, but obtaining some informal evidence of what
happens in such cases is nonetheless of interest, and the results are reported
in rows 9-12 of Table 1. The size results are somewhat worse than in the
earlier two cases for larger $n=1000$, but still they are satisfactory, with
some over-sizing, not in excess of half the nominal size.

\begin{table}[ptbh]
\caption{Monte Carlo size of test of continuity, Setting A}%
\label{tab:addlabel}%
\centering
\par%
\begin{tabular}
[c]{r|rrr|rrr}\hline
$\delta$ & \multicolumn{3}{|c|}{2} & \multicolumn{3}{|c}{$n^{-1/4}\sqrt{10}/4
$}\\
$n\backslash s$ & 0.1 & 0.05 & 0.01 & 0.1 & 0.05 & 0.01\\\hline
100 & 0.1195 & 0.0737 & 0.0204 & 0.152 & 0.0843 & 0.0177\\
250 & 0.0832 & 0.0477 & 0.0122 & 0.1404 & 0.0775 & 0.0162\\
500 & 0.0897 & 0.0408 & 0.0076 & 0.1318 & 0.0684 & 0.0135\\
1000 & 0.105 & 0.0491 & 0.0109 & 0.1312 & 0.0662 & 0.0135\\\hline
$\delta$ & \multicolumn{3}{|c|}{$n^{-1/2}\sqrt{10}/4$} &
\multicolumn{3}{|c}{0}\\
$n\backslash s$ & 0.1 & 0.05 & 0.01 & 0.1 & 0.05 & 0.01\\\hline
100 & 0.1508 & 0.0867 & 0.0165 & 0.1485 & 0.0837 & 0.0177\\
250 & 0.1347 & 0.072 & 0.0144 & 0.1394 & 0.0745 & 0.0147\\
500 & 0.1263 & 0.0653 & 0.0147 & 0.1237 & 0.0633 & 0.0141\\
1000 & 0.1444 & 0.0749 & 0.0154 & 0.14 & 0.0737 & 0.0164\\\hline
\end{tabular}
\end{table}

Tables 2 and 3 report Monte-Carlo power results for the test of continuity for the
nominal size of test $s$ in jump settings B and C, respectively. Power results
naturally are affected by the size of $\delta$, and four sets of $\delta$ have
been tried. For the first three sets, we use values tried in Hansen (2000),
$\delta=0.25, 0.5, 1$, for $n=100$ and let it diminish according to Assumption
\textbf{J} with $\varphi=1/4$. For the fourth set, we fix $\delta=2$ across
$n$. As expected, power improves as $\delta$ gets larger and as $n$ increases.
Power is better in setting C ($Q_{i}\neq X_{i}$) than setting B
($Q_{i}=X_{i}$), which reflects the larger departure of C from A, compared to
that of B. Even in setting B, the reported power results are promising, with
the power being practically 1 for $\delta=2$ with $n=250, 500 $.

\begin{table}[ptbh]
	\caption{Monte Carlo power of test of continuity, Setting B}%
	\label{tab:addlabel}%
	\centering
	\begin{tabular}
		[c]{crr|rrr}\hline
		$\delta$ & $\delta$ & $n \backslash s$ & 0.1 & 0.05 & 0.01\\\hline
		$n^{-1/4}\sqrt{10}/4$ & 0.25 & 100 & 0.1313 & 0.0661 & 0.0134\\
		& 0.1988 & 250 & 0.1205 & 0.0564 & 0.0097\\
		& 0.1672 & 500 & 0.1151 & 0.0574 & 0.0089\\\hline
		$n^{-1/4}\sqrt{10}/2$ & 0.5 & 100 & 0.1525 & 0.0726 & 0.013\\
		& 0.3976 & 250 & 0.1502 & 0.068 & 0.0098\\
		& 0.3344 & 500 & 0.1656 & 0.0787 & 0.0117\\\hline
		$n^{-1/4}\sqrt{10}$ & 1 & 100 & 0.3282 & 0.1918 & 0.0365\\
		& 0.7953 & 250 & 0.4684 & 0.3028 & 0.0623\\
		& 0.6687 & 500 & 0.637 & 0.4797 & 0.1685\\\hline
		\text{fixed} & 2 & 100 & 0.9471 & 0.8854 & 0.6293\\
		& 2 & 250 & 1 & 0.9997 & 0.9986\\
		& 2 & 500 & 1 & 1 & 1\\\hline
	\end{tabular}
\end{table}

\begin{table}[ptbh]
\caption{Monte Carlo power of test of continuity, Setting C}%
\label{tab:addlabel}%
\centering
\begin{tabular}
[c]{crr|rrr}\hline
$\delta$ & $\delta$ & $n \backslash s$ & 0.1 & 0.05 & 0.01\\\hline
$n^{-1/4}\sqrt{10}/4$ & 0.25 & 100 & 0.3756 & 0.2452 & 0.0635\\
& 0.1988 & 250 & 0.4014 & 0.2535 & 0.069\\
& 0.1672 & 500 & 0.4531 & 0.2783 & 0.089\\\hline
$n^{-1/4}\sqrt{10}/2$ & 0.5 & 100 & 0.5779 & 0.4076 & 0.1365\\
& 0.3976 & 250 & 0.7116 & 0.54 & 0.2212\\
& 0.3344 & 500 & 0.8516 & 0.7071 & 0.3729\\\hline
$n^{-1/4}\sqrt{10}$ & 1 & 100 & 0.9638 & 0.9194 & 0.709\\
& 0.7953 & 250 & 0.9978 & 0.9939 & 0.9546\\
& 0.6687 & 500 & 1 & 0.9998 & 0.9988\\\hline
\text{fixed} & 2 & 100 & 1 & 1 & 0.9999\\
& 2 & 250 & 1 & 1 & 1\\
& 2 & 500 & 1 & 1 & 1\\\hline
\end{tabular}
\end{table}

\section{\textbf{EMPIRICAL APPLICATION: GROWTH AND DEBT}}\label{sec:emp}

Reinhart and Rogoff $(2010)$ suggest that above some threshold, the higher debt-to-GDP ratio is related to
a lower GDP growth rate, reporting 90$\%$ as their estimate for the threshold. There have been many studies that investigate the Reinhart-Rogoff hypothesis with the threshold regression models; see Hansen $(2017)$ for references on earlier studies that utilize  discontinuous threshold regression models. Hansen $(2017)$ fitted a kink threshold model to a time series of US annual data and Hidalgo \textit{et al.} (2019) applied their robust inference procedure that is valid for both kink and jump design to Sweden, UK, and Australia data as well as US data used in Hansen (2017). Hansen (2017) mentions that \textquotedblleft one
could imagine testing the assumption of continuity within the threshold model
class. This is a difficult problem, one to which we are unaware of a solution,
and therefore is not pursued in this paper.\textquotedblright As we have
developed testing procedures for continuity in this paper, we follow up on
Hansen's $(2017)$ investigation and present complementary analysis to Hidalgo \textit{et al.} (2019). 

Hansen $(2017)$ used long-span US annual data (1792-2009, $n$=218) on real GDP growth rate in year $t$
($y_{t}$) and debt-to-GDP ratio of the previous year ($q_{t}$) and reported the following estimated
equation with standard errors in parentheses:
\[
\widehat{y}_{t}=\underset{(0.69)}{3.78}+\underset{(0.09)}{0.28}y_{t-1}
+\left\{
\begin{array}
[c]{rr}%
\underset{(0.026)}{0.033}(q_{t}-43.8),\quad\text{if } q_{t} \leq
\underset{(12.1)}{43.8} & \\
-\underset{(0.048)}{0.067}(q_{t}-43.8),\quad\text{if } q_{t} > \underset
{(12.1)}{43.8} &
\end{array}
\right.
\]
We carried out our test of continuity given in Section
\ref{sec:test_stat_cont} with 10,000 bootstraps and obtained $p$-value of
0.029, hence reject the null of continuity at 5$\%$ nominal level. This result is in line with Hansen (2017) that reported $p$-value of 0.15 for the test of the presence of a kink threshold effect. We remark that Hidalgo \textit{et al.} (2019) obtained $p$-value of 0.047 for the test of the presence of threshold effect using Hansen (1996)'s test without imposing the kink model and rejected the null of no threshold effect at 5 $\%$ nominal level. 

The fitted jump model is given by:
\[
\widehat{y}_{t}=\left\{
\begin{array}
[c]{ll}%
\underset{(0.87)}{4.82}-\underset{(0.16)}{0.052}y_{t-1}-\underset
{(0.049)}{0.114}q_{t}, & \text{if }q_{t}\leq17.2\\
\underset{(0.74)}{2.78}+\underset{(0.082)}{0.49}y_{t-1}-\underset
{(0.012)}{0.017}q_{t}, & \text{if }q_{t}> 17.2
\end{array}
\right.
\]
Lower regime contains 99 observations and upper regime contains 109 observations. Hidalgo \textit{et al.} $(2019)$ obtained grid bootstrap confidence intervals for $\tau_{0}$ that are (10.8, 38.6) for 90$\%$ confidence level and
(10.5, 39) for 95$\%$ confidence level. These confidence intervals do not
contain the CLSE $\widetilde{\tau}=43.8$, which is not surprising as the null of continuity is rejected in our test.

Hidalgo \textit{et al.} (2019) also conducts similar analysis with  Sweden data for the period spanning 1881-2009 ($n=129$). The $p$-value for
Hansen (1996)'s test of presence of threshold effect  is reported to be 0.048. Applying our
continuity tests based on 10,000 bootstraps yield $p$-value of 0.091. The
estimated jump model is:
\[
\widehat{y}_{t}=\left\{
\begin{array}
[c]{rr}%
\underset{(2.17)}{1.12}-\underset{(0.24)}{0.2}y_{t-1}+\underset{(0.11)}%
{0.13}q_{t},\quad\text{if } q_{t}\leq{21.3} & \\
\underset{(0.58)}{1.86}+\underset{(0.11)}{0.48}y_{t-1}-\underset
{(0.0082)}{0.004}q_{t},\quad\text{if } q_{t}>21.3 &
\end{array}
\right.
\]
The number of observations of the lower regime is 61, and the upper regime has 68 observations.

The grid bootstrap confidence intervals for $\tau_{0}$ obtained in Hidalgo \textit{et al.} $(2019)$ were (15.3, $\infty$)
and (16.4, $\infty$) for 95$\%$ and 90$\%$ confidence levels. This is in line with our finding that the confidence interval
for $\tau_{0}$ tends to become much wider as the model becomes a kink model,
as reflected by the cube-root convergence rate.

The coefficients of debt-to-GDP ratio were also not significant in
the estimated kink model, which need to be read with caution in the light of
the continuity test:
\[
\widehat{y}_{t}=\underset{(0.58)}{2.89}+\underset{(0.13)}{0.048}%
y_{t-1}+\left\{
\begin{array}
[c]{rr}%
\underset{(0.3)}{0.24}(q_{t}-15.5),\quad\text{if } q_{t}\leq\underset
{(5.75)}{15.5} & \\
-\underset{(0.014)}{0.0008}(q_{t}-15.5),\quad\text{if } q_{t}> \underset
{(5.75)}{15.5} &
\end{array}
\right.
\]
whereby the lower regime had 15 observations and the upper regime contained
114 observations. Note that CLSE $\widetilde{\tau}=15.5$ is contained in the confidence interval.

We conclude that there is substantial heterogeneity across countries in the
relationship between the GDP growth and the debt-to-GDP ratio, not only in the
values of model parameters but also in the type of suitable models.
\footnote{Figures 1-6 in the Appendix present scatterplots of residuals
from autoregression of $y_{t}$ on $y_{t-1}$ against $q_{t}$ for the two
countries, highlighting the importance of deploying the aforementioned tests
in practice. Often neither the economic model nor data plots can tell us much
about the true specification, and one should not expect to be able to spot the
presence of discontinuity from visual inspection of data plots, let alone
discern kink from jump. See Section C of Appendix for some further
discussion.}

\section{\textbf{CONCLUSION}}\label{sec:conclusion}

This paper has developed the continuity test that concerns an interesting hypothesis involving both the regression coefficients and the threshold. The continuity test is complementary to the robust inference presented in Hidalgo \textit{et al.} (2019). The robust inference
concerns inference for each type of parameter separately.

There are several interesting future research topics. First, we have considered the continuity of mean regression function. However, the same issue of continuity also arises in the quantile regression with a threshold. As the continuity of quantile function is not guided by the economic theory, it would be useful to develop a data-driven method for detecting discontinuity of quantile function. Another direction could be to study the high-dimensional model with a threshold. This model has been considered in Lee \textit{et al.} (2016, 2018).  Finally, it would be interesting to find an estimator that matches with the minimax lower bound in Proposition \ref{Lower Bound}. 

\appendix

\section{\textbf{PROOFS OF MAIN THEOREMS}}

\subsection{\textbf{Proof of Proposition \ref{Consistency}}}
Hidalgo \textit{et al.} (2019) considers the case where the threshold is fixed over the sample size, namely, $\varphi=0$. We generalize this result to the diminishing threshold, $0<\varphi<1/2$. Without loss of generality, we may assume that $\tau_0=0$. Let $\overline{\psi}:=\psi-\psi_0$ for any parameter $\psi$ and $\mathbb{I}_i(a;b)=\mathbb{I}\{a<Q_i<b\}$. Denote $v:=\beta+\delta$.

We derive the convergence rate of the LSE, that is,
we show that%
\[
\left(  \sqrt{n}\left(  \widehat{\delta}_{1}-\delta_{01}\right)  ,\sqrt
{n}\left(  \widehat{\delta}_{3}-\delta_{03}\right)  ,n^{\frac{1-2\varphi}{3}%
}\left(  \widehat{\tau}-\tau_{0}\right)  \right)  =O_{p}\left(  1\right)
.
\]
Note that we can write
\[
{{\mathbb{S}}}_{n}\left(  \theta\right)  -{{\mathbb{S}}}_{n}\left(  \theta
_{0}\right)  =\mathbb{A}_{n1}\left(  \theta\right) 
+\mathbb{A}_{n2}\left(  \theta\right) +\mathbb{A}_{n3}\left(
\theta\right)  +\mathbb{B}_{n1}\left(  \theta\right) 
+\mathbb{B}_{n2}\left(  \theta\right)+\mathbb{B}_{n3}\left(
\theta\right)  \text{,}%
\]
where 
\begin{align*}
\mathbb{A}_{n1}\left(  \theta\right)    & =\bar{v}^{\prime}\frac{1}{n}\sum_{i=1}^{n}X_{i}X_{i}^{\prime}\mathbb{I}_{i}\left(\tau\right)
\bar{v}; \quad\quad \mathbb{A}_{n2}(\theta)=\bar{\beta}^{\prime}\frac{1}{n}\sum_{i=1}^{n}X_{i}X_{i}^{\prime}\mathbb{I}_{i}\left(-\infty;0\right)
\bar{\beta}\\
\mathbb{A}_{n3}\left(  \theta\right)    & =\left(\bar{\beta}+\delta_{0}\right)^{\prime}\frac{1}{n}%
\sum_{i=1}^{n}X_{i}X_{i}^{\prime}\mathbb{I}_{i}\left(  0;\tau\right)
\left(\bar{\beta}+\delta_{0}\right)\\
\mathbb{B}_{n1}\left(  \theta\right)    & =\bar{v}^{\prime}\frac{2}{n}\sum_{i=1}^{n}{X}_{i}{U}_{i}\mathbb{I}_{i}\left(  \tau\right)  ; \quad\quad \mathbb{B}_{n2}(\theta)=\bar{\beta}\frac{2}{n}\sum_{i=1}^{n}{X}_{i}{U}_{i}\mathbb{I}_i\left(-\infty;0\right)\\
\mathbb{B}_{n3}\left(  \theta\right)    & =\left(\bar{\beta}+\delta_{0}\right)^{\prime}\frac{2}{n}%
\sum_{i=1}^{n}X_{i}U_{i}\mathbb{I}_{i}\left(  0;\tau\right)
\text{,}%
\end{align*}
and $\tau>0$. The case where $\tau<0$ can be handled similarly. We follow the approach taken in Hidalgo \textit{et al.} (2019, Proposition 1), for which we
need to verify that for any $\epsilon>0$, there exist $C>0$, $\eta>0$ and $n_0$ such that for all $n>n_0$,
\begin{equation}
\Pr\left\{  \inf_{\frac{C}{n^{1/2}}<\left\Vert \overline{\upsilon}\right\Vert
,\left\Vert \overline{\beta}\right\Vert <\eta;\frac{C}{n^{\left(
1-2\varphi\right)  /3}}<\left\Vert \overline{\tau}\right\Vert <\eta}%
\sum_{\ell=1}^{3}  \left(\mathbb{A}_{n\ell}\left(  \theta\right)  
+\mathbb{B}_{n\ell}\left(  \theta\right)\right)  \leq0\right\}  <\epsilon\text{.}\label{dimconv}
\end{equation}
Note the change of the lower bound for $\bar
{\tau}$ from $n^{1/3}$ to $n^{\left(  1-2\varphi\right)  /3}$. To prove $(\ref{dimconv})$, it suffices to show that for each $\ell=1,2,3$,  \begin{equation}
    \Pr\left\{  \inf_{\frac{C}{n^{1/2}}<\left\Vert \overline{\upsilon}\right\Vert
,\left\Vert \overline{\beta}\right\Vert <\eta;\frac{C}{n^{\left(
1-2\varphi\right)  /3}}<\left\Vert \overline{\tau}\right\Vert <\eta}%
E\left(  \mathbb{A}_{n\ell}\left(  \theta\right)  \right)/2
+\left(\mathbb{A}_{n\ell}\left(  \theta\right)-E\left(  \mathbb{A}_{n\ell}\left(  \theta\right)  \right) \right) \leq0\right\}  <\epsilon\text{,}
\end{equation} 
and
\begin{equation}
   \Pr\left\{  \inf_{\frac{C}{n^{1/2}}<\left\Vert \overline{\upsilon}\right\Vert
,\left\Vert \overline{\beta}\right\Vert <\eta;\frac{C}{n^{\left(
1-2\varphi\right)  /3}}<\left\Vert \overline{\tau}\right\Vert <\eta}%
E\left(  \mathbb{A}_{n\ell}\left(  \theta\right)  \right)/2 + \mathbb{B}_{n\ell}\left(\theta\right)  \leq0\right\}  <\epsilon\text{.}\label{eq:convrate}
\end{equation} Notice that the only difference from the proof of Proposition 1 in Hidalgo \textit{et al.} (2019) due to the assumption of $\delta_{0}=d_0\cdot
n^{-\varphi}$ lies in the case $\ell=3$. Therefore, it is sufficient to handle the
contribution from $\mathbb{A}_{n3}\left(  \theta\right)  $ and $\mathbb{B}%
_{n3}\left(  \theta\right)  .$ Since $E\left(X_iX_i'\mathbb{I}_i(0;\tau)\right)$ is positive definite, we may consider 
\begin{align*}
    \widetilde{\mathbb{A}}_{n3}(\theta) = (\beta_{3}-\beta_{30}+\delta_{30})^2\frac{1}{n}\sum_{i=1}^{n}Q_i^2\mathbb{I}_i(0;\tau); \quad \widetilde{\mathbb{B}}_{n3}(\theta)=(\beta_{3}-\beta_{30}+\delta_{30})\frac{2}{n}\sum_{i=1}^{n}Q_{i}U_{i}\mathbb{I}_i(0;\tau).
\end{align*}Accordingly, we decompose the parameter space over which the infimum is taken as
\begin{align*}
\Xi_{k}  &=\left\{  \theta:\frac{C}{n^{1/2}}<\rVert\bar{\upsilon} \rVert, \rVert \bar{\beta} \rVert<\eta,\frac{C2^{k-1}}{n^{(1-2\varphi)/3}}%
<\overline{\tau}<\frac{C2^{k}}{n^{\left(  1-2\varphi\right)  /3}}%
\right\}  ; \;k=1,...,\log_{2}\left(\frac{\eta}{C}n^{(1-2\varphi)/3}\right).
\end{align*}
Recall that we have assumed that $\tau\geq0$, as the case $\tau\leq0$ follows similarly.

Also recall that we impose that $\delta_{30}%
=d_{3}\cdot n^{-\varphi}\ $. Choose a positive real number $C_1$ such that $E\left(Q_{i}^{2}\mathbb{I}_{i}\left(  0;\xi\right)\right)  \geq
C_{1}\xi^{3}$ and $\left\vert d_{3}\right\vert >C_{1}>0\ $. Then, we have
\begin{align*}
& \Pr\left\{  \inf_{\Xi_{k}  }E\left(  \widetilde{\mathbb{A}}_{n3}\left(  \theta\right)
\right)/2  +\widetilde{\mathbb{B}}_{n3}\left(  \theta\right)  \leq0\right\}
\\
& \leq\Pr\left\{  \inf_{\Xi_{k} }\left\vert
d_{3}\right\vert n^{-\varphi}E\left(  Q_{i}^{2}\mathbb{I}_{i}\left(
0;\tau\right)  \right)  \leq\sup_{\Xi_{k} }\left\Vert
\frac{4}{n}\sum_{i=1}^{n}Q_{i}U_{i}\mathbb{I}_{i}\left(
0;\tau\right)  \right\Vert \right\}  \\
& \leq\Pr\left\{  \frac{C_{1}^2C^3}{32n^{\left(  1-2\varphi\right)/2  }%
}2^{3k  }\leq\sup_{\Xi_{k} }\left\Vert
\frac{1}{n^{1/2}}\sum_{i=1}^{n}
Q_{i}U_{i}\mathbb{I}_{i}\left(
0;\tau\right)  \right\Vert \right\}  \\
& \leq (32C^{-3/2}C_{1}^{-2}C_2)2^{-3k/2}\text{,}%
\end{align*}
by Lemma \ref{lem:maxineq2} and Markov's inequality where $C_2$ is a constant in Lemma \ref{lem:maxineq2}. Letting $C$ be sufficiently large, we obtain the inequality (\ref{eq:convrate}) from the summability of $2^{-3k/2}$. The remaining steps to obtain the convergence rate of $n^{\left(  1-2\varphi\right)  /3}$ are
identical to Hidalgo \textit{et al.} (2019, Proposition 1 and Theorem 1). \hfill$\blacksquare$

\subsection{\textbf{Proof of Proposition \ref{Lower Bound}}}
$\left.  {}\right.  $
The proof for the lower bound in Proposition \ref{Lower Bound} relies on Le Cam's method (Le Cam, 1973). Before proceeding to the proof, we collect some notations and basic properties of divergence measures. Let $\mathbb{P}$, $\mathbb{Q}$ be any probability measures on the measurable space $(\mathcal{X}, \mathcal{A})$, where $\mathcal{A}$ is a $\sigma$-field on $\mathcal{X}$. Then the total variation distance between $\mathbb{P}$ and $\mathbb{Q}$ is defined as $d_{TV}(\mathbb{P},\mathbb{Q}):=\sup_{A\in\mathcal{A}}|\mathbb{P}(A)-\mathbb{Q}(A)|$ and Kullback-Leibler(KL) divergence from $\mathbb{P}$ to $\mathbb{Q}$ is $d_{KL}(\mathbb{P},\mathbb{Q})= \int \log{\frac{d\mathbb{P}}{d\mathbb{Q}}}d\mathbb{P}$ if $\mathbb{P}$ is absolutely continuous with respect to $\mathbb{Q}$, or $+\infty$, otherwise. It is known that for all probability measures $\mathbb{P}$ and $\mathbb{Q}$,
\begin{equation}\label{Pinsker}
    d_{TV}(\mathbb{P},\mathbb{Q}) \leq \sqrt{\frac{1}{2}d_{KL}(\mathbb{P}, \mathbb{Q})},
\end{equation}
which is called Pinsker's inequality. Finally, consider a regression model, $Y = g(X)+U$, where $U\sim\mathcal{N}(0,\sigma^2(X))$ given $X$. We write $\mathbb{P}_{g}$ for the joint distribution of $(Y,X)$. Assume that $\sigma^2(X)\geq\underline{\sigma}^2>0$. Then $d_{KL}(\mathbb{P}_{g_0},\mathbb{P}_{g_1})\leq \frac{1}{2\underline{\sigma}^2}\rVert g_1 - g_0 \rVert_{\ell_2(\mathbb{P}_X)}^2$.

We state a version of Le Cam's method from Yu (1996). Let $\mathcal{P}$ be a class of probability measures. Let $X_1, X_2, \cdots, X_n$ be random variables sampled from $\mathbb{P}\in\mathcal{P}$ in \textit{i.i.d.} manner and $\mathbb{P}^n$ denote the corresponding product measure. Define a function $\theta$ which maps a probability measure in $\mathcal{P}$ into the metric space $\Theta$ with a metric $\rho$. We write $\widehat{\theta}=\widehat{\theta}(X_1, X_2, \cdots, X_n)$ for an estimator of $\theta$. For any probability measure $\mathbb{P}_0, \mathbb{P}_1\in\mathcal{P}$, the $\rho$-minimax risk is lower bounded as follows: 

\begin{equation}\label{Le Cam}
    \inf_{\widehat{\theta}}\sup_{\mathbb{P}\in\mathcal{P}}E_{\mathbb{P}}\bigl(\rho(\theta(\mathbb{P}),\widehat{\theta}(X_1, \cdots, X_n))\bigr) \geq \rho(\theta(\mathbb{P}_0),\theta(\mathbb{P}_1))\frac{1-d_{TV}(\mathbb{P}_0^n,\mathbb{P}_1^n)}{2},
\end{equation}
where the inifimum is taken over all estimators $\widehat{\theta}$. 

Note that  $d_{KL}(\mathbb{P}_0^n,\mathbb{P}_1^n) = n d_{KL}(\mathbb{P}_0,\mathbb{P}_1)$. Combining (\ref{Le Cam}) with Pinsker's inequality, it is straightforward to see that the minimax risk is lower bounded as follows:
$\left.  {}\right.  $
\begin{lemma}\label{KL LeCam}Let $\{\mathbb{P}_{0,n}\}_{n\in\mathbb{N}}$ and $\{\mathbb{P}_{1,n}\}_{n\in\mathbb{N}}$ be any two sequences of probability measures in $\mathcal{P}$. Let $\{\mathbb{P}_{0,n}\}_{n\in\mathbb{N}}$ and $\{\mathbb{P}_{1,n}\}_{n\in\mathbb{N}}$ satisfy 
\begin{align*}
    d_{KL}(\mathbb{P}_{0,n}, \mathbb{P}_{1,n})\leq \frac{1}{2n}
\end{align*}
for all $n\in\mathbb{N}$. Then,
\begin{align*}
    \inf_{\widehat{\theta}}\sup_{\mathbb{P}\in\mathcal{P}}E_{\mathbb{P}}\bigl(\rho(\theta(\mathbb{P}),\widehat{\theta}(X_1, \cdots, X_n))\bigr) \geq \frac{1}{4}\rho(\theta(\mathbb{P}_{0,n}),\theta(\mathbb{P}_{1,n}))
\end{align*}
for all $n\in\mathbb{N}$, where the infimum is taken over all estimators $\widehat{\theta}$.
\end{lemma}
We prove Proposition \ref{Lower Bound} with this lemma. Since we are considering \textit{i.i.d.} sampling, we drop the subscript $i$ of random variables. Let $\mathbb{P}_{(\alpha,\gamma)}$ denote the joint distribution of $(Y, X)$ where $Y=X'(\tau)\alpha+U$, and $U\sim\mathcal{N}(0,\sigma^2(Q))$ given $X$. Let $\xi=\inf \mathbb{T}$, $\beta = 0\in\mathbb{R}^d$, $\delta = (-\kappa\xi ,0,\cdots,0,\kappa)$ and $\alpha = (\beta',\delta')'$. First, we consider the case that $n\geq \frac{3\underline{\sigma}^2}{\overline{f}\kappa^2\eta^3}$. Let $\tau_0=\xi$ and $\tau_{1,n} = \xi + \bigl(\frac{3\underline{\sigma}^2}{n\overline{f}\kappa^2} \bigr)^{1/3}$, then $\tau_0, \tau_{1,n}\in \mathbb{T}$ for all $n$. We can obtain the following inequality under this choice of parameter sequences,
\begin{align*}
    d_{KL}\bigl(\mathbb{P}_{(\alpha,\tau_0)}, \mathbb{P}_{(\alpha,\tau_{1,n})}\bigr) &\leq \frac{1}{2\underline{\sigma}^2}\rVert \kappa(Q-\tau_0) \mathbb{I}\{\tau_0<Q\leq\tau_{1,n}\} \rVert_{\ell_2(\mathbb{P}_X)}^2 \\
    &= \frac{\kappa^2}{2\underline{\sigma}^2}\int_{\tau_0}^{\tau_{1,n}}(q-\tau_0)^2f(q)dq \leq \frac{\overline{f}\kappa^2}{6\underline{\sigma}^2}(\tau_{1,n}-\tau_0)^3 = \frac{1}{2n}.
\end{align*}
Applying Lemma \ref{KL LeCam} with $\{\mathbb{P}_{(\alpha,\tau_0)}\}_{n\in\mathbb{N}}$ and $\{\mathbb{P}_{(\alpha,\tau_{1,n})}\}_{n\in\mathbb{N}}$, we get the desired result.

Next, assume that $n<\frac{3\underline{\sigma}^2}{\overline{f}\kappa^2\eta^3}$. In this case, we let $\tau_1 = \sup\mathbb{T}$. Then,
\begin{align*}
    d_{KL}(\mathbb{P}_{(\alpha,\tau_0)}, \mathbb{P}_{(\alpha, \tau_1)}) &\leq \frac{\kappa^2}{2\underline{\sigma}^2}\int_{\tau_0}^{\tau_1}(q-\tau_0)^2f(q)dq\\
    & \leq \frac{\overline{f}\kappa^2\eta^3}{6\underline{\sigma}^2}\leq \frac{1}{2n}.
    \end{align*}
Therefore, the minimax risk is lower bounded by $\frac{\eta}{4}$ as desired. \hfill$\blacksquare$

\subsection{\textbf{Proof of Theorem \ref{Th:AD_cube}}}
Theorem \ref{Th:AD_cube} is parallel to the Hidalgo \textit{et al.} (2019, Theorem 1). Therefore, we briefly review the proof and emphasize the difference caused by the diminishing threshold assumption.

Observing the continuity of ``argmin" function and the convergence rates in Proposition \ref{Consistency}, we only need to consider the weak limit of
\begin{align}
    \mathbb{G}_{n}(h, g) = n\left(\mathbb{S}_n\left(\alpha_0+\frac{h}{n^{1/2}},\frac{g}{n^{(1-2\varphi)/3}}\right)-\mathbb{S}_n\left(\alpha_0,0\right)\right),
\end{align}
where $\tau_0$ is assumed to be $0$. Note that 
\begin{equation*}
    \sup_{\rVert h \rVert , |g| \leq C}\left| \mathbb{G}_{n}(h,g)-\widetilde{\mathbb{G}}_n(h,g)\right|=o_p(1),
\end{equation*}
where 
\begin{align*}
    \widetilde{\mathbb{G}}_{n}\left(h,g\right)  = &\left\{ h'\frac{1}{n}\sum_{i=1}^n \mathbf{X}_{i}\mathbf{X}_{i}^{\prime}h-h'\frac{2}{n^{1/2}}\sum_{i=1}^{n}\mathbf{X}_{i}U_{i}\right\} \\
    &+\delta_{30}\left\{\delta_{30}\sum_{i=1}^{n}Q_i^2\mathbb{I}_{i}\left(0;\frac{g}{n^{(1-2\varphi)/3}}\right)-2\sum_{i=1}^{n}Q_{i}U_{i}\mathbb{I}_{i}\left(0;\frac{g}{n^{(1-2\varphi)/3}}\right) \right\} \\
    = &:\widetilde{\mathbb{G}}_{n}^{1}(h)+\widetilde{\mathbb{G}}_{n}^{2}(g).
\end{align*}
Comparing to Proposition 1 of Hidalgo \textit{et al.} (2019), it suffices to examine $\widetilde{\mathbb{G}}_n^2(g)$. Note that the first term of $\widetilde{\mathbb{G}}_n^2(g)$ uniformly converges to $3^{-1}d_{30}^2f(0)|g|^3$ due to the Lemma $\ref{lem:maxineq2}$.

Next, we show that the second term, $-2\delta_{30}\sum_{i=1}^{n}Q_{i}U_{i}\mathbb{I}_{i}\left(0;\frac{g}{n^{(1-2\varphi)/3}} \right)$, weakly converges to $2d_{30}\sqrt{\frac{\sigma^2(0)f(0)}{3}}W(g^3)$. Let $Z_{ni} = n^{(1-2\varphi)/2}Q_{i}U_{i}\mathbb{I}_{i}\left(0;\frac{g}{n^{(1-2\varphi)/3}}\right)$. Note that
\begin{align*}
    \frac{1}{n}\sum_{i=1}^n Z_{ni}^2 \overset{p}{\longrightarrow}\frac{\sigma^2(0)f(0)}{3}|g|^3\text{.}
\end{align*}
Covariances are calculated to be
\begin{align*}
   n^{1-2\varphi}E\left(Q_i^2U_i^2\mathbb{I}_i\left\{\frac{g_1}{n^{(1-2\varphi)/3}},\frac{g_2}{n^{(1-2\varphi)/3}}\right\}\right)=\frac{\sigma^2(0)f(0)}{3}\left(g_2^3-g_1^3\right) +o(1) \text{,}
\end{align*}
where $g_2>g_1$, other cases can be treated similarly. Therefore, the second term of $\widetilde{\mathbb{G}}_n^2(g)$ converges to $2d_{30}\sqrt{\frac{\sigma^2(0)f(0)}{3}}W(g^3)$ from the martingale CLT. 

Similar analysis on the covariance shows the aysmptotic independence between $\widetilde{\mathbb{G}}_n^1(h)$ and $\widetilde{\mathbb{G}}_n^2(g)$ Remaining details are identical to Hidalgo \textit{et al.} (2019).
\hfill$\blacksquare$

\subsection{\textbf{Proof of Theorem \ref{Th:Q_n}}}

$\left.  {}\right.  $
From the convergence rate in Proposition \ref{Consistency}, we examine the weak limit of %
\[
\mathbb{G}_n(h,g) = n\left\{\mathbb{S}_n\left(\alpha_0+\frac{h}{n^{1/2}},\tau_0+\frac{g}{n^{(1-2\varphi)/3}}\right)-\mathbb{S}_n\left(\alpha_0,\tau_0\right) \right\},
\]
for $0\leq\varphi<1/2$. From the proof of Theorem \ref{Th:AD_cube},
\[
\sup_{\rVert h \rVert, |g| \leq C}\left|\mathbb{G}_n
(h,g) - \widetilde{\mathbb{G}}_{n}^{1}(h) - \widetilde{\mathbb{G}}_{n}^{2}(g)\right|   =  o_p(1) \text{,}
\] where
\begin{align*}
\widetilde{\mathbb{G}}_{n}^{1}\left(  h\right)   & =\left\{  h^{\prime}%
\frac{1}{n}\sum_{i=1}^{n}\mathbf{X}_{i}\mathbf{X}_{i}^{\prime}h-2h^{\prime}\frac
{1}{n^{1/2}}\sum_{i=1}^{n}\mathbf{X}_{i}U_{i}\right\} \\
\widetilde{\mathbb{G}}_{n}^{2}\left(  g\right)   & =\delta_{30}\left\{
\delta_{30}\sum_{i=1}^{n}Q_{i}^{2}\mathbb{I}_{i}\left(  0;\frac{g}{n^{1/3}%
}\right)  -2\sum_{i=1}^{n}Q_{i}U_{i}\mathbb{I}_{i}\left(  0;\frac
{g}{n^{(1-2\varphi)/3}}\right)  \right\}
\end{align*}
and $\mathbb{\widetilde{G}}_{n}^{j}\left(  \cdot\right)  ,j=1,2,$ are mutually
independent. Therefore, for the unconstrained estimator
$\widehat{\alpha}$ and $\widehat{\tau},$ we could write%
\[
n\left(  \mathbb{S}_{n}\left(  \widehat{\alpha},\widehat{\tau}\right)
-\mathbb{S}_{n}\left(  \alpha_{0},\tau_{0}\right)  \right)  =\min
_{h}\mathbb{\widetilde{G}}_{n}^{1}\left(  h\right)  +\min_{g}%
\mathbb{\widetilde{G}}_{n}^{2}\left(  g\right)  +o_{p}\left(  1\right).
\]
 Similarly, for the constrained estimator $\widetilde{\alpha}$ and
$\widetilde{\tau}$ we can write, see e.g. Chan and Tsay $(1998)$ or Hansen
$(2017)$, that%
\[
n\left(  \mathbb{S}_{n}\left(  \widetilde{\alpha},\widetilde{\tau}\right)
-\mathbb{S}_{n}\left(  \alpha_{0},\tau_{0}\right)  \right)  =\min_{\ell
}\mathbb{H}_{n}\left(  \ell\right)  +o_{p}\left(  1\right)  ,
\]
where
\[
\mathbb{H}_{n}\left(  \ell\right)  =\ell^{\prime}\frac{1}{n}\sum_{i=1}%
^{n}\mathbf{\bar{X}}_{i}\mathbf{\bar{X}}_{i}^{\prime}\ell-2\ell^{\prime}%
\frac{1}{n^{1/2}}\sum_{i=1}^{n}\mathbf{\bar{X}}_{i}U_{i}%
\]
and $\mathbf{\bar{X}}_{i}=\left(  X_{i}^{\prime},\left(  Q_{i}-\tau
_{0}\right)  \mathbb{I}_{i}(\tau_0)  ,-\left(  \beta_{30}%
+\delta_{30}\right)  \mathbb{I}_{i}(\tau_0) \right)  ^{\prime
}=R\mathbf{X}_{i}$. Note that $\mathbb{H}_{n}\left(  \ell\right)
+\mathbb{\widetilde{G}}_{n}^{1}\left(  h\right)  $ converges weakly as a
function of $h$ and $\ell$ since both $\ell^{\prime}\frac{1}{n}\sum_{i=1}%
^{n}\mathbf{\bar{X}}_{i}\mathbf{\bar{X}}_{i}^{\prime}\ell$ and $h^{\prime
}\frac{1}{n}\sum_{i=1}^{n}\mathbf{X}_{i}\mathbf{X}_{i}^{\prime}h$ converges uniformly
in probability by ULLN and $\ell^{\prime}\frac{1}{\sqrt{n}}\sum_{i=1}%
^{n}\mathbf{\bar{X}}_{i}U_{i}+h^{\prime}\frac{1}{\sqrt{n}}\sum
_{i=1}^{n}\mathbf{X}_{i}U_{i}$ converges weakly by the linearity,
the CLT and Cramer-Rao device. The weak convergence of $\mathbb{\widetilde{G}%
}_{n}^{2}\left(  g\right)  $ to  $2d_{30}\sqrt{\frac{\sigma^{2}\left(  \tau_{0}\right)
f\left(  \tau_{0}\right)  }{3}}W\left(  g^{3}\right)  +\frac{d_{30}%
^{2}}{3}f\left(  \tau_{0}\right)  \left\vert g\right\vert ^{3}$ where the gaussian process $W$ is defined in Theorem \ref{Th:AD_cube} and its asymptotic independence from
$\mathbb{\widetilde{G}}_{n}^{1}\left(  h\right)  $ are given in the proof of Theorem \ref{Th:AD_cube}. By the same argument it is asymptotically independent of
$\mathbb{H}_{n}\left(  \ell\right)  $. To sum up, let
\begin{align*}
\mathbb{K}\left(  h,g,\ell\right)   & =\ell^{\prime}E\mathbf{\bar{X}}%
_{i}\mathbf{\bar{X}}_{i}\ell+h^{\prime}E\mathbf{X}_{i}\mathbf{X}_{i}h-2\left(
\ell^{\prime}R+h^{\prime}\right)  B\\
& +\left(  2d_{30}\sqrt{\frac{\sigma^{2}\left(  \tau_{0}\right)
f\left(  \tau_{0}\right)  }{3}}W\left(  g^{3}\right)  +\frac{d_{30}%
^{2}}{3}f\left(  \tau_{0}\right)  \left\vert g\right\vert ^{3}\right)
\text{,}%
\end{align*}
where $B$ is a $\mathcal{N}\left(  0,\Omega\right)  $ and independent of the gaussian
process $W$,  and let%
\[
\mathbb{K}_{n}\left(  h,g,\ell\right)  =\mathbb{\widetilde{G}}_{n}^{1}\left(
h\right)  +\mathbb{\widetilde{G}}_{n}^{2}\left(  g\right)  +\mathbb{H}%
_{n}\left(  \ell\right)  .
\]
Then, it follows from the preceding discussion that
\[
\mathbb{K}_{n}\left(  h,g,\ell\right)  \Rightarrow\mathbb{K}\left(
h,g,\ell\right)  \text{.}%
\]
Furthermore,%
\begin{align*}
& n\left(  \mathbb{S}_{n}\left(  \widetilde{\alpha},\widetilde{\tau}\right)
-\mathbb{S}_{n}\left(  \widehat{\alpha},\widehat{\tau}\right)  \right) \\
& =\min_{\ell
:g=0,h=0}\mathbb{K}_{n}\left(  h,g,\ell\right) -\min_{h:g=0,\ell=0}\mathbb{K}_{n}\left(  h,g,\ell\right)  -\min
_{g:h=0,\ell=0}\mathbb{K}_{n}\left(  h,g,\ell\right)   +o_{p}\left(  1\right) \\
& \overset{d}{\longrightarrow}\min_{\ell:g=0,h=0}\mathbb{K}\left(  h,g,\ell\right) -\min_{h:g=0,\ell=0}\mathbb{K}\left(
h,g,\ell\right)  -\min_{g:h=0,\ell=0}\mathbb{K}\left(  h,g,\ell\right)
 \text{,}%
\end{align*}
due to the continuous mapping theorem as the (constrained) minimum is a
continuous operator and the fact that $\mathbb{\widetilde{G}}_{n}^{1}\left(
h\right)  ,\mathbb{\widetilde{G}}_{n}^{2}\left(  g\right)  \ $and
$\mathbb{H}_{n}\left(  \ell\right)  $ are zero at the origin. Certainly this
limit is $O_{p}\left(  1\right)  $ and does not degenerate since
$\mathbb{\widetilde{G}}_{n}^{2}\left(  g\right)  $ is asymptotically
independent of the other terms. The convergence of $\mathbb{\widehat{S}}_{n}$
is straightforward by standard algebra and the ULLN and CLT and thus details
are omitted.%
\[
\]

$\hfill\blacksquare$

\subsection{Proof of Theorem \ref{Th:QBoot}}

$\left.  {}\right.  $

Recalling the meaning of the superscript \textquotedblleft$^{\ast}$\textquotedblright{\Large \ } in section \ref{sec:btrp c_test}, we begin by observing the 
consistency and rate of convergence of $\left(\widehat{\alpha}^{\ast}, \widehat{\tau}^{\ast}\right)$. 

\begin{proposition}
\label{ConsistencyBoot}Suppose that Assumptions \ref{A:Z}, \ref{A:Q} and \textbf{C} hold. Let $\delta_0=d_0\cdot n^{-\varphi}$. If $0\leq\varphi<1/2$, \newline%
$\left(  \mathbf{a}\right)  $ 
\[
\widehat{\alpha}^{\ast}-\widetilde{\alpha}=O_{p^{\ast}}\left(  n^{-1/2}\right)
\ \ \ \text{and \ \ }\widehat{\tau}^{\ast}-\tau_{0}=O_{p^{\ast}}\left(
n^{-(1-2\varphi)/3}\right)  \text{,}%
\]
\newline
\noindent $\left(  \mathbf{b}\right)  $ $\widehat{\alpha}^{\ast}$ and $\widehat{\tau}^{\ast}$ are asymptotically independent and (in probability)
\begin{align*}
& n^{1/2}(\widehat{\alpha}^{\ast}-\widetilde{\alpha})\overset{d^{\ast}}{\longrightarrow}%
\mathcal{N}\left(  0,M^{-1}\Omega M^{-1}\right) \\
& n^{(1-2\varphi)/3}(\widehat{\tau}^{\ast}-\tau_{0})\overset{d^{\ast}}{\longrightarrow}%
\underset{g\in\mathbb{R}}{\operatorname{argmax}}\left(2d_{30}\sqrt
{\frac{\sigma^{2}\left(  \tau_{0}\right)  f\left(  \tau_{0}\right)  }{3}%
}W\left(  g^{3}\right)  +\frac{d_{30}^{2}}{3}f\left(  \tau_{0}\right)
\left\vert g\right\vert ^{3}\right)\text{.}
\end{align*}
\end{proposition}

Proposition\ \ref{ConsistencyBoot} (a) is similar to Proposition 5 (a) of Hidalgo \textit{et al.} (2019). The only difference is that the centering term of the resampling scheme is $\left(\widetilde{\alpha}, \widetilde{\tau} \right)$ instead of $\left(\widehat{\alpha}, \widehat{\tau} \right)$. Following the proof of Hidalgo \textit{et al.} (2019, Theorem 3 (a)), we obtain Proposition \ref{ConsistencyBoot} (b). Theorem \ref{Th:QBoot} is a direct consequence of Proposition \ref{ConsistencyBoot} and the same argument as the proof of Theorem \ref{Th:Q_n}.\hfill$\blacksquare$

\section{\textbf{AUXILIARY LEMMA}}

Refer to Hidalgo \textit{et al.} (2019) for the proofs of the lemmas in this section. For
$j=1$ or $2,$ let
\begin{align*}
J_{n}\left(  \tau,\tau^{\prime}\right)   & =\frac{1}{n^{1/2}}\sum
_{i=1}^{n}U_{i}X_{i}\mathbb{I}_{i}\left(  \tau;\tau^{\prime
}\right) \\
J_{1n}\left(  \tau,\tau^{\prime}\right)   & =\frac{1}{n^{1/2}}\sum
_{i=1}^{n}U_{i}\left\vert Q_{i}-\tau\right\vert ^{j}\mathbb{I}%
_{i}\left(  \tau;\tau^{\prime}\right) \\
J_{2n}\left(  \tau\right)   & =\frac{1}{n^{1/2}}\sum_{i=1}^{n}\left\{
\left\vert Q_{i}-\tau_{0}\right\vert ^{j}\mathbb{I}_{i}\left(  \tau
_{0};\tau\right)  -E\left\vert Q_{i}-\tau_{0}\right\vert ^{j}%
\mathbb{I}_{i}\left(  \tau_{0};\tau\right)  \right\}
\end{align*}
and for some sequence $\left\{  Z_{i}\right\}  _{i=1}^{n}$,
\[
J_{3n}\left(  \tau\right)  =\frac{1}{n^{1/2}}\sum_{i=1}^{n}\left(
Z_{i}\mathbb{I}_{i}\left(  \tau_{0};\tau\right)  -EZ_{i}\mathbb{I}%
_{i}\left(  \tau_{0};\tau\right)  \right)  \text{.}%
\]

\begin{lemma}
\label{lem:maxineq2}Suppose Assumptions 1 and 2 hold for the sequence
$\left\{  X_{i},U_{i}\right\}  _{i=1}^{n}$. In addition, for
$J_{3n}\left(  \tau\right)  ,$ assume that $\left\{  Z_{i},Q_{i}\right\}
_{i=1}^{n}$ be a sequence of strictly stationary, ergodic, and $\rho$ -mixing
with $\sum_{m=1}^{\infty}\rho_{m}^{1/2}<\infty$, $E\left\vert Z_{i}\right\vert
^{4}<\infty$ and, for all $\tau\in\mathbb{T}$,$\ ${{\ \textrm{$E$}}}$\left(
\left\vert Z_{i}\right\vert ^{4}|Q_{i}=\tau\right)  <C<\infty$. Then, there
exists $n_{0}<\infty$ such that for all $\tau^{\prime}\ $in a neighbourhood
of $\tau_{0}\ $and for all $n>n_{0}$ and $\epsilon\geq n_{0}^{-1}$,
\begin{align*}
&\left(  \mathbf{a}\right)  \text{~~}E\sup_{\tau^{\prime}<\tau
<\tau^{\prime}+\epsilon}\left\vert J_{n}\left(  \tau^{\prime}%
,\tau\right)  \right\vert  \leq C\epsilon^{1/2}\nonumber\\
&\left(  \mathbf{b}\right)  \text{~~}E\sup_{\tau^{\prime}<\tau
<\tau^{\prime}+\epsilon}\left\vert J_{1n}\left(  \tau^{\prime}%
,\tau\right)  \right\vert  \leq C\epsilon^{1/2}\left(  \epsilon+\left\vert
\tau_{0}-\tau^{\prime}\right\vert \right)  ^{j}\\
&\left(  \mathbf{c}\right)\text{~~}E\sup_{\tau_{0}<\tau<\tau
_{0}+\epsilon}\left\vert J_{2n}\left(  \tau\right)  \right\vert   \leq
C\epsilon^{j+1/2}\\
&\left(  \mathbf{d}\right)\text{~~}E\sup_{\tau_{0}<\tau<\tau_{0}+\epsilon}\left\vert J_{3n}\left(  \tau\right)  \right\vert  \leq
C\epsilon^{1/2}\text{,}
\end{align*}
where $j=1$ or $2$.
\end{lemma}

\renewcommand\thesection{C-\arabic{section}} \setcounter{section}{0}

\section{Figures for Empirical Application in Section 7}

Figures 1-6 below are scatter plots of residuals from fitting AR(1) model on
$y_{t}$, plotted against $q_{t}$, superimposed with the estimated jump and
kink models for the US and Sweden.

As is made clear by these figures, one cannot expect to spot presence of
discontinuity visually by examining the scatter plots, let alone discern if
the kink or jump models better fits the data. To illustrate this point, in
Figures 5-6 we present the same scatter plots based on simulated data that
were generated from the estimated jump equations of the two countries, which
used $y_{0}, q_{t}$ from the data and $u_{t}\sim \mathcal{N}(0, s^{2})$, with sample
variance of the residuals $s^{2}$, to reconstruct $y_{t}$. They are both
superimposed with the jump equation that is the true data generating process for the simulated data.

This lack of visual guidance is indeed why the testing procedures of presence
of threshold effect of e.g. Andrews (1993), Hansen (1996), Lee \textit{et al}.
(2011), and our continuity testing procedure of the current paper are very
much needed, and should be deployed in data analysis. \vskip.3cm

\begin{figure}[ptb]
\begin{center}
\includegraphics[width=7in]{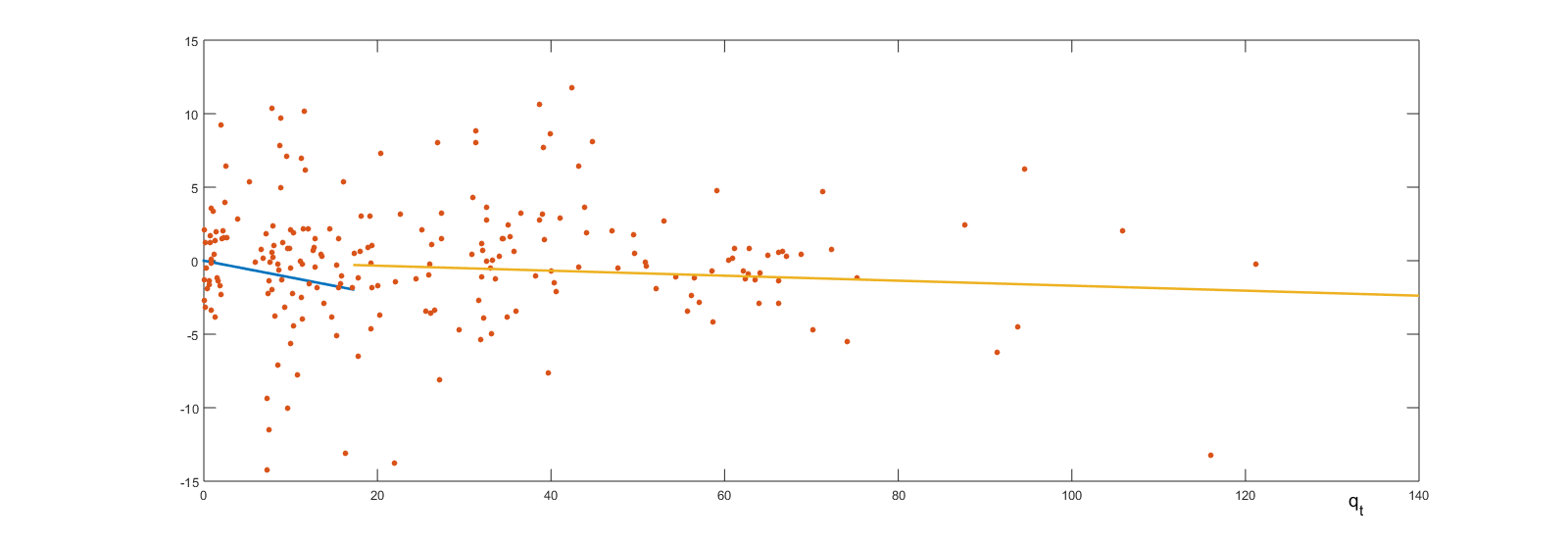}
\end{center}
\caption{Scatter plot of AR(1) residuals of $y_{t}$ against $q_{t}$ and
estimated jump equation, US}%
\end{figure}

\begin{figure}[ptb]
\begin{center}
\includegraphics[width=7in]{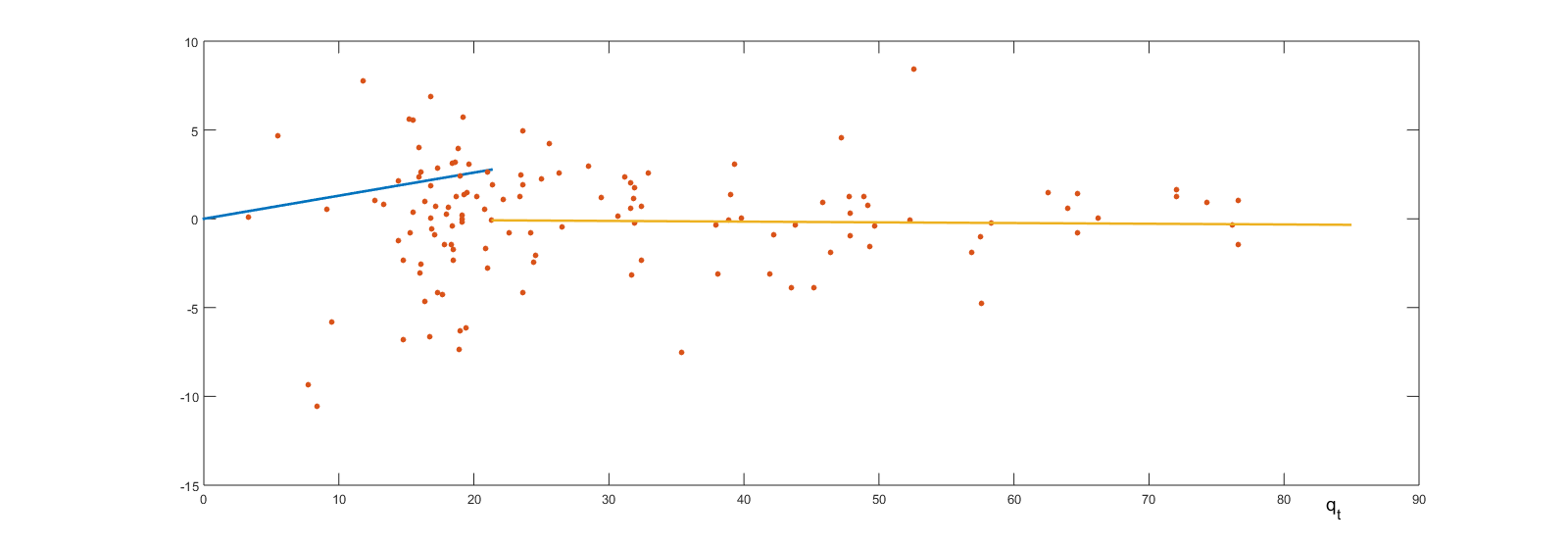}
\end{center}
\caption{Scatter plot of AR(1) residuals of $y_{t}$ against $q_{t}$ and
estimated jump equation, Sweden}%
\end{figure}

\begin{figure}[ptb]
\begin{center}
\includegraphics[width=7in]{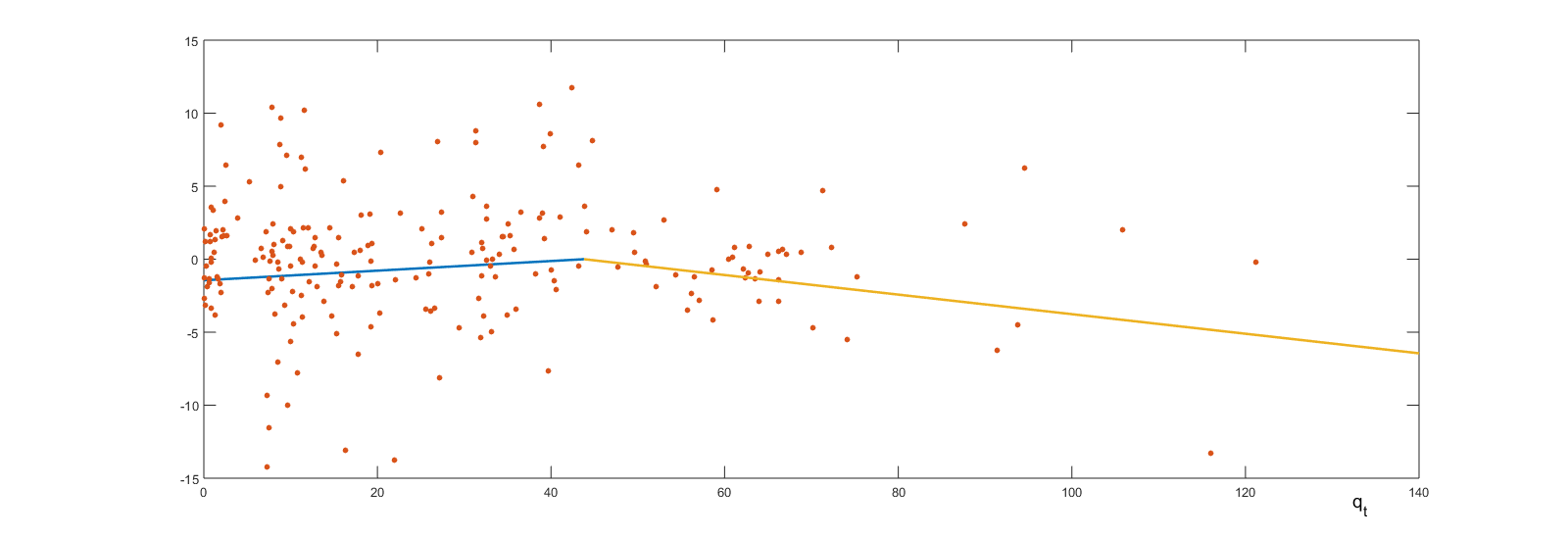}
\end{center}
\caption{Scatter plot of AR(1) residuals of $y_{t}$ against $q_{t}$ and
estimated kink equation, US}%
\end{figure}

\begin{figure}[ptb]
\begin{center}
\includegraphics[width=7in]{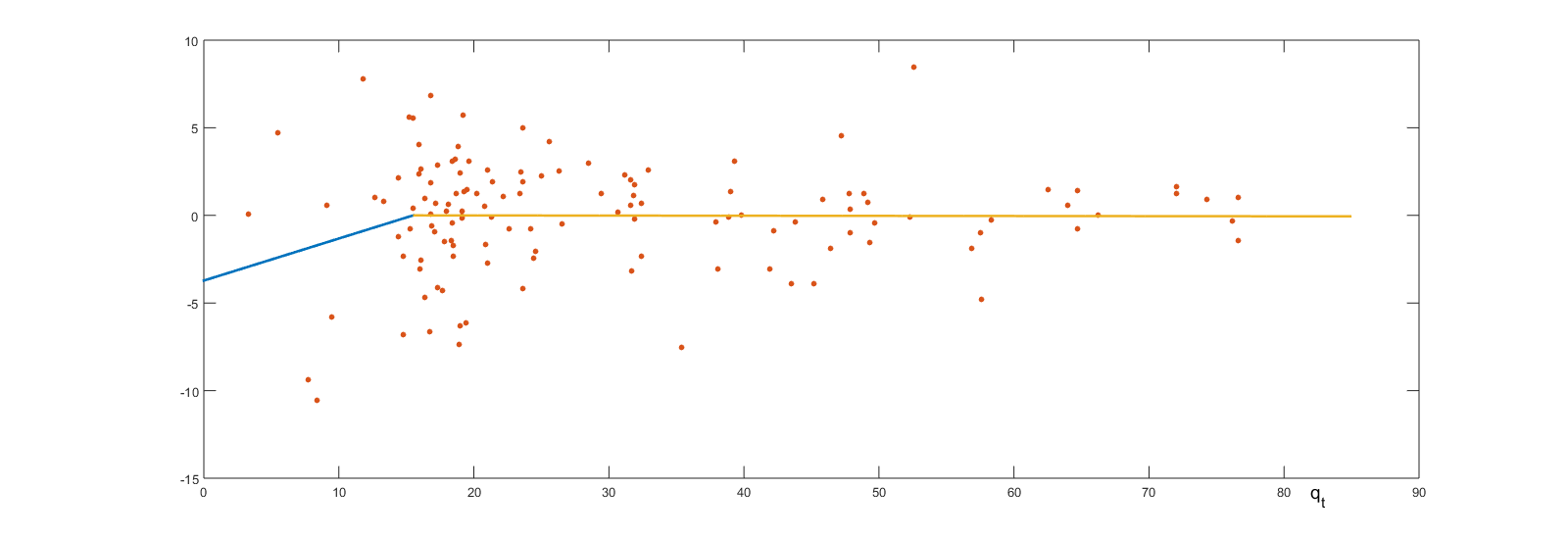}
\end{center}
\caption{Scatter plot of AR(1) residuals of $y_{t}$ against $q_{t}$ and
estimated kink equation, Sweden}%
\end{figure}

\begin{figure}[ptb]
\begin{center}
\includegraphics[width=7in]{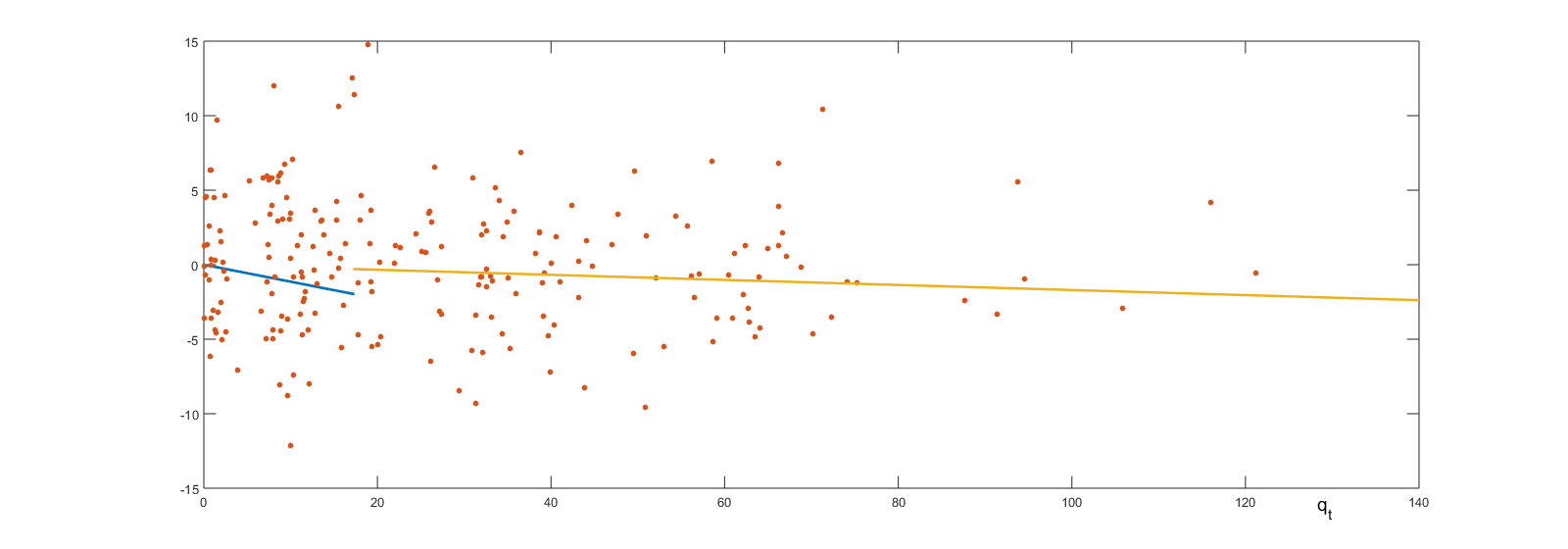}
\end{center}
\caption{Scatter plot of AR(1) residuals of $y_{t}$ against $q_{t}$ from
reconstructed data for US, and true jump equation}%
\end{figure}

\begin{figure}[ptb]
\begin{center}
\includegraphics[width=7in]{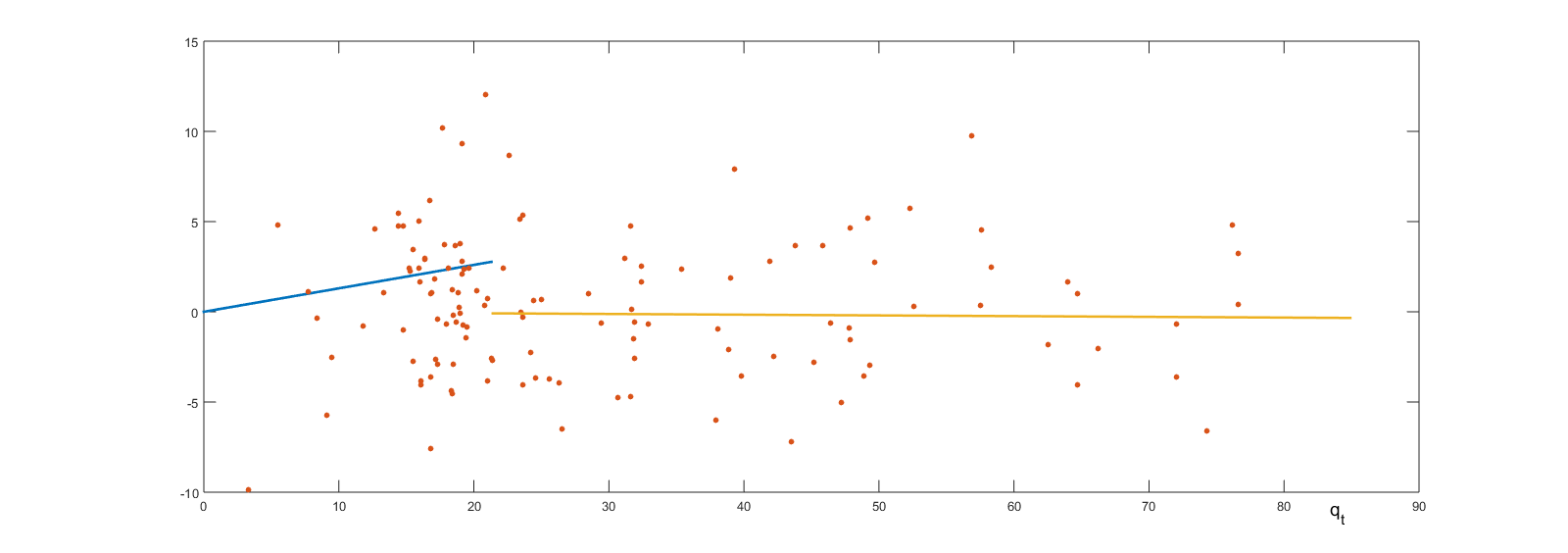}
\end{center}
\caption{Scatter plot of AR(1) residuals of $y_{t}$ against $q_{t}$ from
reconstructed data for Sweden, and true jump equation}%
\end{figure}

\end{document}